\documentclass[sigconf]{acmart}
\usepackage{amsfonts}
\usepackage{balance} 
\def\BibTeX{{\rm B\kern-.05em{\sc i\kern-.025em b}\kern-.08emT\kern-.1667em\lower.7ex\hbox{E}\kern-.125emX}}

\copyrightyear{2020}
\acmYear{2020}
\setcopyright{acmcopyright}\acmConference[SIGIR '20]{Proceedings of the 43rd International ACM SIGIR Conference on Research and Development in Information Retrieval}{July 25--30, 2020}{Virtual Event, China}
\acmBooktitle{Proceedings of the 43rd International ACM SIGIR Conference on Research and Development in Information Retrieval (SIGIR '20), July 25--30, 2020, Virtual Event, China}
\acmPrice{15.00}
\acmDOI{10.1145/3397271.3401426}
\acmISBN{978-1-4503-8016-4/20/07}
\usepackage{graphicx}
\usepackage{subfigure}
\usepackage{multirow}
\begin{document}

%
% The "title" command has an optional parameter, allowing the author to define a "short title" to be used in page headers.
\title{A Heterogeneous Information Network based Cross Domain Insurance Recommendation System for Cold Start Users}
\author{Ye Bi, Liqiang Song, Mengqiu Yao, Zhenyu Wu, Jianming Wang, Jing Xiao}
\email{magicyebi@163.com, {songliqiang537, yaomengqiu621, wuzhenyu447, wangjianming888, xiaojing661}@pingan.com.cn}
\affiliation{%
\institution{Ping An Technology Shenzhen Co., Ltd}
}
\begin{abstract}
Internet is changing the world, adapting to the trend of internet sales will bring revenue to traditional insurance companies. Online insurance is still in its early stages of development, where cold start problem (prospective customer) is one of the greatest challenges. In traditional e-commerce field, several cross-domain recommendation (CDR) methods have been studied to infer preferences of cold start users based on their preferences in other domains. However, these CDR methods couldn't be applied to insurance domain directly due to the domain's specific properties. In this paper, we propose a novel framework called a Heterogeneous information network based Cross Domain Insurance Recommendation (HCDIR) system for cold start users. Specifically, we first try to learn more effective user and item latent features in both source and target domains. In source domain, we employ gated recurrent unit (GRU) to module users' dynamic interests. In target domain, given the complexity of insurance products and the data sparsity problem, we construct an insurance heterogeneous information network (IHIN) based on data from PingAn Jinguanjia, the IHIN connects users, agents, insurance products and insurance product properties together, giving us richer information. Then we employ three-level (relational, node, and semantic) attention aggregations to get user and insurance product representations. After obtaining latent features of overlapping users, a feature mapping between the two domains is learned by multi-layer perceptron (MLP). We apply HCDIR on Jinguanjia dataset, and show HCDIR significantly outperforms the state-of-the-art solutions.
\end{abstract}



\maketitle

\section{Introduction}

\begin{figure}[!t]
\centering
\subfigure[Home Page]{
\begin{minipage}[t]{0.3\linewidth}
\centering
\label{figure22}
\includegraphics[width=1.05in]{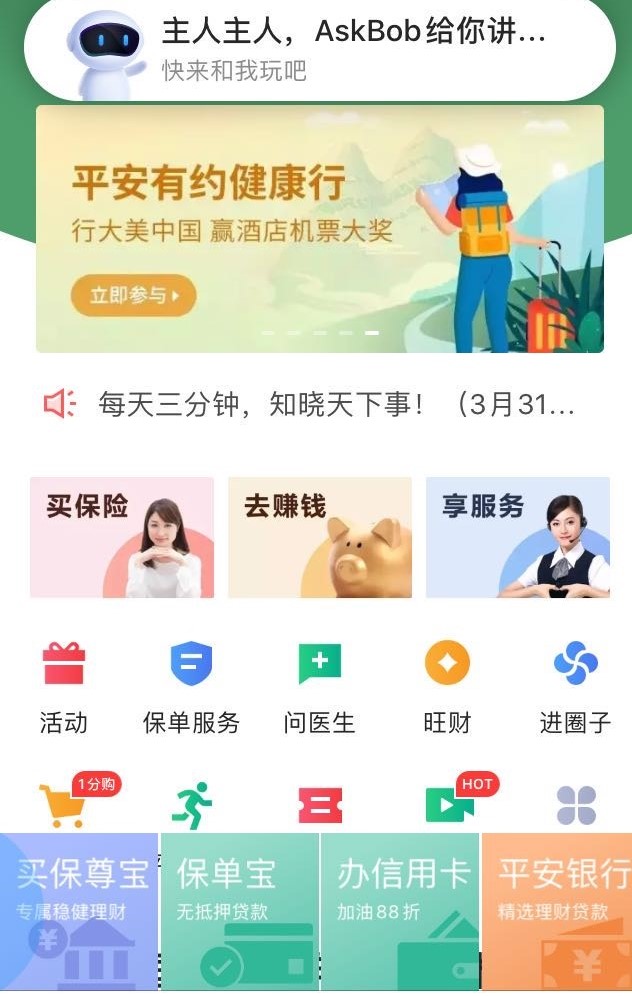}
\end{minipage}
}
\subfigure[Nonfinanacial Domain]{
\begin{minipage}[t]{0.3\linewidth}
\centering
\label{figure21}
\includegraphics[width=1.05in]{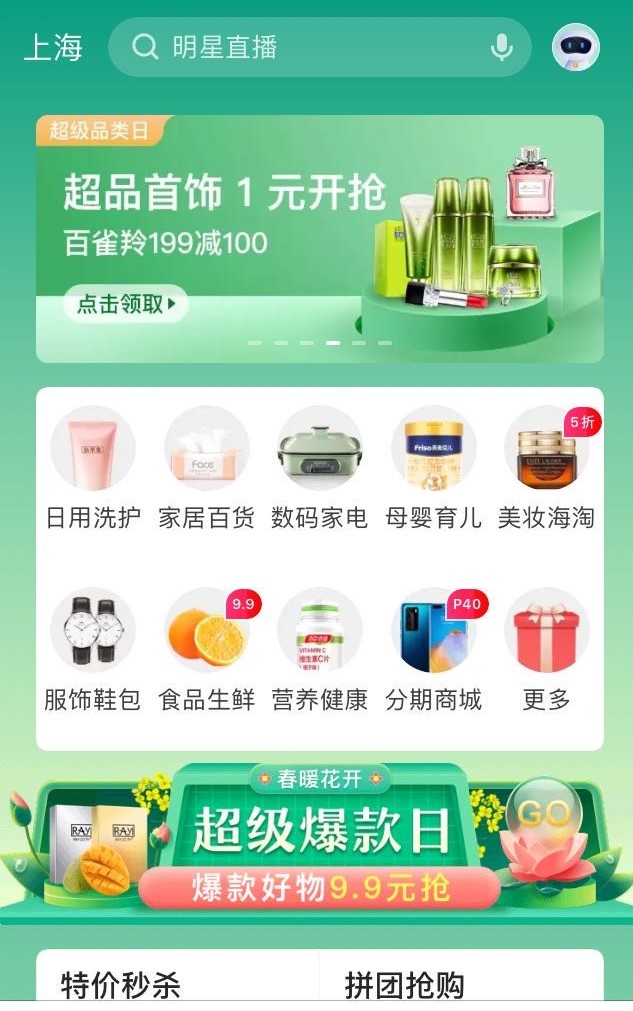}
\end{minipage}
}
\subfigure[Insurance Domain]{
\begin{minipage}[t]{0.3\linewidth}
\centering
\label{figure22}
\includegraphics[width=1.05in]{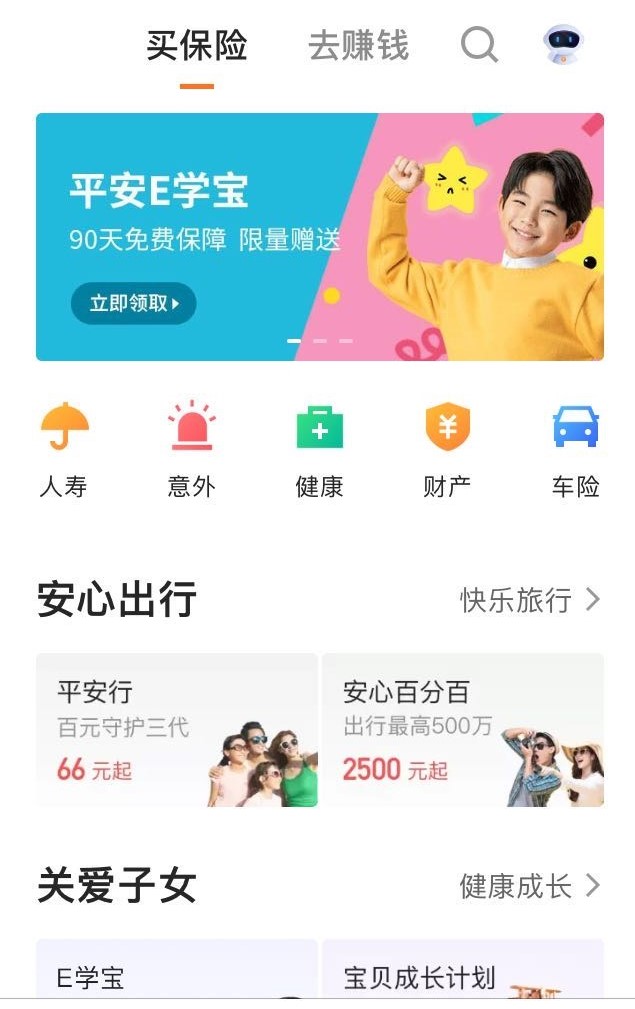}
\end{minipage}
}
\caption{Online shopping on Jinguanjia. (a) is homepage. (b) is nonfinancial domain, providing daily necessities. (c) is insurance domain, providing various insurance products.}
\label{SIGIR2020_Industry_JGJ}
\end{figure}

Internet is changing the world, every segment of the economy is experiencing dramatic change and is having to respond to shifts in the value chain, enhanced consumer power, and altered competitive cycles. Internet insurance adapted to the trend of economic boom in internet age for two main sectors. For supply side, internet insurance overcomes the limitations of live sales and geography, increasing the customer base. For demand side, internet sales are more acceptable to young people, who are the main consumers of insurance products. Adapting to the trend of internet sales will bring revenue to traditional insurance companies.

Internet insurance is still in its early stages of development, where cold start problem (prospective customer) is one of the greatest challenges. For example, PingAn Jinguanjia, one of the most popular comprehensive applications (App) in China, which boasts more than 100 million registered users, has nearly $90\%$ cold start users in insurance domain (i.e. these registered users didn't buy any insurance products). This situation resulted from many reasons. First, insurance policies are so complex that ordinary users are relatively lack of knowledge to understand them. Besides, insurance products are typically bought to be used for a long time period (e.g. one year in car insurance). Attracting prospective customers plays a critical role in buildup of the competitive edge for traditional insurance company. Under this circumstances, our motivation for creating an online insurance recommendation system stems from providing personalize recommendations for prospective customers, and then building customer loyalty. To our knowledge, there are not many models about recommendation systems in insurance domain, some includes \cite{DBLP:conf/sac/RokachSSCS13,DBLP:conf/recsys/QaziFMF17,DBLP:conf/icmcs/LiuZKZZ019}. However, these methods treat insurance domain and traditional e-commerce equally, neglecting product complexity and data sparsity problem in insurance domain.

In this paper, we focus on PingAn Jinguanjia, one of the most popular comprehensive applications in China. In addition to traditional e-commerce products (defined as nonfinancial products in this paper), e.g. electronics, household supplies, etc., it also provides financial products like insurance products, investment services. Besides, each registered customer would be assigned with an agent, who can help with enquiries, offer recommendations.
As mentioned above, even though Jinguanjia has a big user group, it does not have a greater share of sales in online insurance. In other words, most of the registered users did not buy any insurance products, though they have relatively abundant activities in nonfinancial domain. As a result, we could not get enough information in insurance domain only. Traditional recommendation systems, like collaborate filtering (CF) \cite{DBLP:conf/www/AbdollahiN16}, sequential-based models \cite{DBLP:journals/corr/ChungGCB14}, could not perform effectively in insurance domain, since most of users only have less than 2 interactions in a year. To obtain enough information and get more accurate recommendation, PingAn company tries to use side information form Jinguanjia App (the interaction behaviors form nonfinancial domain), but to little avail.

Cross-domain recommendation (CDR) \cite{DBLP:conf/ijcai/ManSJC17,DBLP:conf/sigir/MaRLCMR19,DBLP:conf/cikm/KangHLY19,DBLP:conf/aaai/FuPWXL19}, which aims to improve the recommendation performance by means of transferring information from the source domain to the target domain, is one of the promising ways to solve data sparsity and cold start problem. These methods assume that there exists overlap in information between users and/or items across different domains, and train a mapping function from the source-domain into the target-domain. So the key factor for CDR method is to learn more comprehensive and accurate user representations in two domain. However, the complexity of the insurance products and the severe data sparsity hinder us from learning user representations in insurance domain as accurate as possible. As a result, we could not apply CDR methods into insurance and nonfinancial domain directly.

In order to help the users understand the complex insurance policies and get user representations as comprehensive and accurate as possible, an insurance heterogeneous information network (IHIN) is constructed according to the data from Jinguanjia App. In IHIN, we define four types of nodes corresponding to users, agents, insurance products and insurance product properties, and six types of edges denoting various types of relations between them. Graph convolutional networks (GCN) \cite{DBLP:conf/nips/HamiltonYL17} and Graph attention networks (GAT) \cite{DBLP:conf/iclr/VelickovicCCRLB18} as powerful deep representation learning method for graph data, has shown superior performance on recommendation. However, these methods apply identical aggregation function on various types of edges, and the number of neighbors grows exponentially as the layers stacked up, which prohibit these methods performing efficiently on HIN. To deal with heterogeneous information, many state-of-the-art models emerge and has been proved to be efficient \cite{DBLP:conf/www/WangJSWYCY19,DBLP:conf/cikm/XuLHLX019,DBLP:conf/esws/SchlichtkrullKB18}. R-GCNs \cite{DBLP:conf/esws/SchlichtkrullKB18} are developed to deal with highly multi-relational data. HAN \cite{DBLP:conf/www/WangJSWYCY19} designs a two level (node-level and semantic-level) attentions to generate node embedding by aggregating features from meta-path based neighbors.

Inspired by these models, we propose a novel framework called a Heterogeneous information network based Cross Domain Insurance Recommendation (HCDIR) system for cold start users. Specifically, we first try to learn more effective user and item latent features in both source and target domains. In source domain, users interactions are rich, we can easily get the consume sequence of users, so we employ gated recurrent unit (GRU) \cite{DBLP:journals/corr/ChungGCB14} to module users' dynamic interests. In target domain, given the complexity of insurance products and the data sparsity problem, we construct an IHIN based on data from Jinguanjia App, the IHIN connects users, agents, insurance products and insurance product properties together, giving us richer underlying information. Then we employ three-level (relational, node, and semantic) attention aggregations to get user and insurance product representations. After obtaining the latent features of the overlapping users, a feature mapping between the two domains is learned by multi-layer perceptron (MLP). 

In summary, our contributions in this paper are as follows:
\begin{itemize}
\item To the best of our knowledge, this is the first work to combine cross-domain mechanism and heterogeneous information network to give personalized recommendations for cold start users in insurance domain.
\item For the complexity of insurance products, we construct a heterogeneous information network, which contains four types of nodes and six types of relations. And we employ three level aggregations over IHIN to learn more effective user and item representations in insurance domain.
\item We conduct experiments on real-world recommendation scenarios, and the results prove the efficacy of HCDIR over several state-of-the-art baselines.
\end{itemize}

\section{Data and Preliminary}
\subsection{Dataset}
\label{SIGIR2020_Industry_data_des}
Our dataset is collected from one of the largest e-commerce platform PingAn Jinguanjia. As shown in Figure \ref{SIGIR2020_Industry_JGJ}, 
Jinguanjia provides not only nonfinancial products (traditional e-commerce products), but also financial products like insurance products, investment services. Besides, each registered customer would be assigned with an agent, who can help with enquiries, offer recommendations. In this paper, we aim at providing recommendations to prospective customers by CDR method in insurance domain, the users we use are overlapping users, who have interactions in both insurance domain and nonfinancial domain. Our dataset is collected within the time period from June 1st 2018 to May 31th 2019, the statistics of which is shown in Table \ref{SIGIR2020_Industry_dataset}.

\begin{table}[!h]
\setlength{\abovecaptionskip}{0cm}
\setlength{\belowcaptionskip}{-0.2cm}
\centering
\footnotesize{
\caption{Statistics of Our dataset.}
\label{SIGIR2020_Industry_dataset}
\begin{tabular}{c|c|c|c}
\toprule[1pt]
\multicolumn{2}{c|}{IS-domain (Target domain)}&\multicolumn{2}{c}{NF-domain (Source domain)	}\\
\hline
$\#$User Nodes                        & 117,613                     & $\#$Users                                       & 117,613   \\
$\#$Item Nodes				        &	42	                    	&	$\#$Items				                        &	19,266		\\
$\#$Agent Nodes    		            &	90,377	                &	$\#$User-Item Interactions				&	1,995,168		\\
$\#$Insurance Property Nodes & 35               &                 &    \\
$\#$User-Iten Relations				&	344,206		&					&		\\
$\#$User-Agent Relations          &	97,343	    &					&		\\
$\#$Item-Property Relations      &	275      	    &					&		\\
\bottomrule[1.0pt]
\end{tabular}
}
\end{table}

\textbf{Nonfinancial Domain.} The nonfinancial domain contains pursue logs of nonfinancial products (daily necessities) including clothes, skincare products, fruits, etc. Each item is associated with a description, illustrating category, function, and so on. Besides, we also have the interaction order of each user.

\textbf{Insurance Domain.} The insurance domain contains short-term insurance (coverage time less than one year) including illness insurances, accident insurances, medical insurances, education insurances and other kinds of insurances. To better learn user representations, we construct an insurance heterogeneous information network, which contains four types of nodes (user (U), agent (A), insurance product (I), insurance property (P)), and six types of relations among them (U$\leftrightarrow$I: \emph{purchase} and \emph{be purchased by}; U$\leftrightarrow$A: \emph{be served by} and \emph{serve}; I$\leftrightarrow$P: \emph{possess} and \emph{be possessed by}). Insurance policies are very complex, even the same product, if two customers are in different age groups, they may pay different price. To better describe insurance products, we choose 35 insurance properties (e.g. price, age limit, coverage time, etc.) that customers care most about as nodes in IHIN.

\subsection{Observations in Real Data}

Is it necessary to design a recommendation system specifically for cold start users in online insurance domain? To answer this, we start by investigating the following questions. 

\begin{figure}[!h]
\setlength{\abovecaptionskip}{-0.2cm}
\setlength{\belowcaptionskip}{-0.2cm}
  \centering
  \includegraphics[scale=0.4]{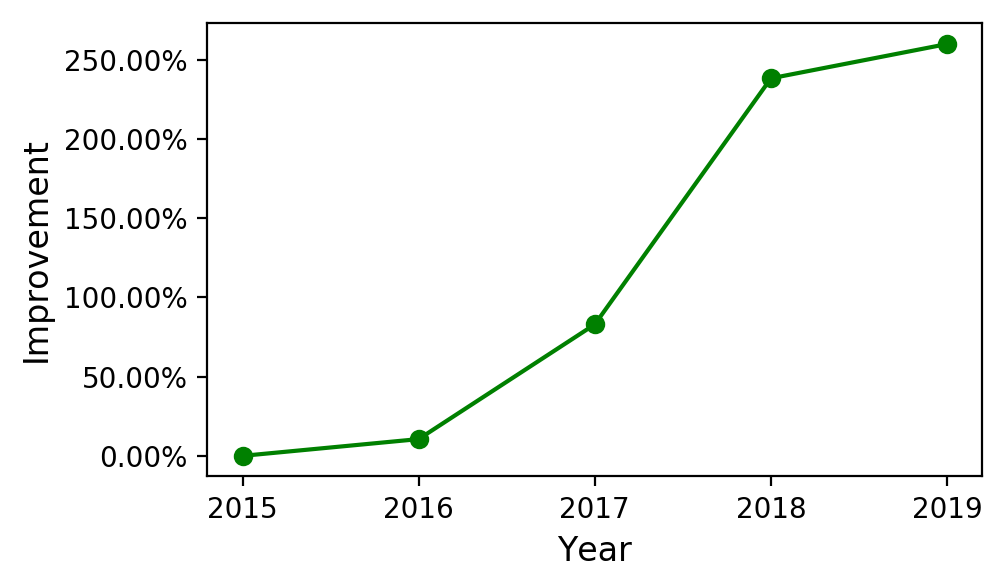}
  \caption{The Number of Insured Improvements w.r.t. 2015 on Jinguanjia App}
  \label{SIGIR2020_Industry_improvement}
\end{figure}

\textbf{Q1. Is online insurance the tendency?} First, we might wonder that if users really buy insurance products online, or they may have been used to buying insurance products in the traditional way. To answer this question, we calculate the number of insured improvements on Ping An Jinguanjia App from 2015 to 2019, which are showed in Figure \ref{SIGIR2020_Industry_improvement}. From the results, we can observe that online insurance experienced explosive growth from 2016 to 2018, the number of insured orders on Jinguanjia jumped by over 2 times in 2018 compared with that in 2015. However, the growth in 2019 entered a bottleneck period, so it is urgent for insurance company to adjust their operation patterns to the internet trend.

\begin{figure}[!h]
\setlength{\abovecaptionskip}{-0.2cm}
\setlength{\belowcaptionskip}{-0.2cm}
  \centering
  \includegraphics[scale=0.4]{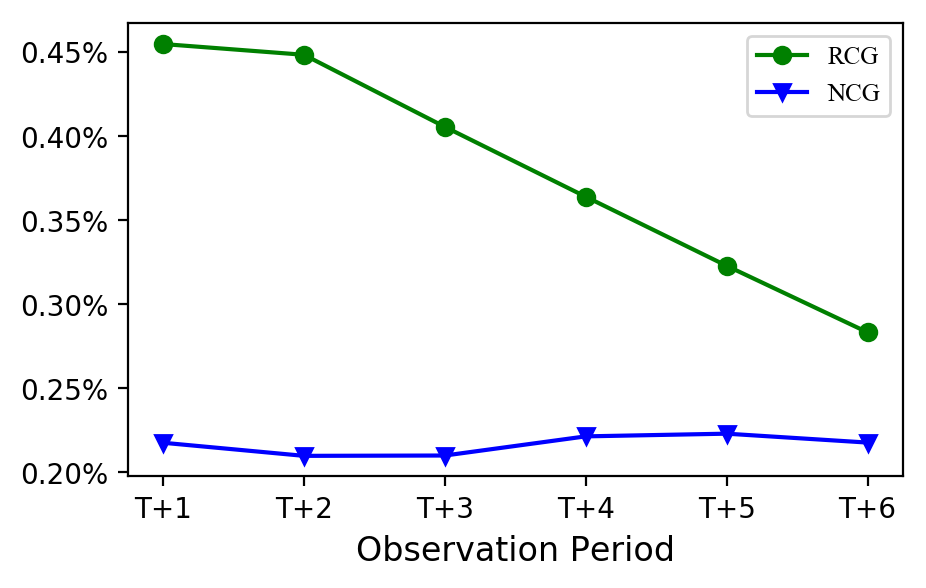}
  \caption{Group-buy-ratio of RCG and NCG.}
  \Description{Group-buy-ratio of RCG and NCG.}
  \label{SIGIR2020_Industry_gbr}
\end{figure}

\textbf{Q2. Do users' behaviours in nonfinancial domain have influence on their behaviors in insurance domain?} In order to investigate the implicit relationships between users' behaviors in insurance domain and nonfinancial domain, we define a metric called group-buy-ratio. For a given group, the group-buy-ratio is defined by the number of people who buy insurance products on Jinguanjia for the first time divided by the total number of people in this group.
\begin{eqnarray*}
\text{group-buy-ratio}=\frac{\text{number of people first buying insurance}}{\text{total number of people in the group}}
\end{eqnarray*}
We first select two groups of customers in Jinguanjia, regular customer group (RCG) includes customers who have bought only some nonfinancial products on Jinguanjia before, new customer group (NCG) concludes new registered customers, i.e. customers who did't buy anything. Then we calculate group-buy-ratio for the two group in six months and summarize them in Figure \ref{SIGIR2020_Industry_gbr}. 

From Figure \ref{SIGIR2020_Industry_gbr}, we can see that group-buy-ratio of RCG is higher than that of NCG. There may be two reasons. Firstly, regular customers in nonfinancial domain might be more willing to trust Jinguanjia, since they have shopping experience in the App. Moreover, as shown in Figure \ref{SIGIR2020_Industry_JGJ}(b), most goods Jinguanjia provides are health products, customers who buy those products may concern more about themselves. So, users' behaviors in nonfinancial domain may help us make recommendations in insurance domain. We note here that group-buy-ratio in each month is new purchases rate in the group. Since insurance products are typically bought to be used for a long time period (e.g. one year for car insurance), the customers may not buy them again in a short time period, so group-buy-ratio of RCG decreases in our statistic period. Even though, group-buy-ratio of RCG is higher than that of NCG.

\textbf{Q3. Are insurance policies really complex?} In traditional e-commerce domain, for example, in clothes, customers only need to see picture to decided weather they need the cloth or not. In insurance domain, understanding items may require a considerable cognitive overload. For example, there are 11 main terms and 30 subsidiary terms in ``PingAn critical illness insurance clause''. The main terms include responsibilities of insurance company, exemption of insured liability, rights and obligations for both policy holder and insurer, etc. The subsidiary terms includes some explanation of  medical term and exception of the insurance. In a word, insurance policies are complex, and the complexity can be summarized as numerous contents and complicated terminology.

\textbf{Q4. Are agents affecting users' buying behaviors in insurance domain?} To investigate whether agents are affecting users' buying decisions, we define ask-buy-ratio, which is the number of customers who buy online insurance after consulting the agent dividing by the total number of customers who have consulted. We first divide the agents according to their communication frequency with customers, and list the ask-buy-ratio of the top $5\%$, top $10\%$, and top $15\%$ communication frequency agents in Table \ref{SIGIR2020_Industry_agent}. From Table \ref{SIGIR2020_Industry_agent}, we can see that different agents have different ask-buy-ratio, ask-buy-ratio of the top $5\%$ communication frequency agents is more than four times that of the top $15\%$ communication frequency agents. This indicates that if a customer is assigned with an agent in the top $5\%$ communication frequency group, he /she may be more likely to buy online insurance products.
\begin{table}[!h]
\setlength{\abovecaptionskip}{-0.06cm}
\setlength{\belowcaptionskip}{-0.2cm}
\centering
\footnotesize{
\caption{Ask-buy-ratio of different agents.}
\label{SIGIR2020_Industry_agent}
\begin{tabular}{c|c|c|c}
\toprule[1pt]
communication frequency order	&	T+1	&	T+2	&	T+3	\\
\hline
top $5\%$  &	4.9784$\%$	&	4.9750$\%$	&	4.7591$\%$	\\
\hline
top $10\%$ &	2.2700$\%$	&	2.3214$\%$	&	2.2698$\%$	\\
\hline
top $15\%$  &	1.0184$\%$	&	1.0635$\%$	&	1.0145$\%$	\\
\bottomrule[1.0pt]
\end{tabular}
}
\end{table}

To sum up, we have following findings.
\begin{itemize}
\item Online insurance is becoming more and more popular, thought it is in its growth bottleneck. A special recommendation system for online insurance domain is in demand.
\item Users' behaviors in nonfinancial domain have influence on their behaviors in insurance domain. Users who have shopping experiences in nonfinancial domain are more willing to trust Jinguanjia, and more likely to buy insurance products.
\item Insurance policies are too complex to understand, the traditional randomly initialized method could not give the accurate item representations.
\item Different agents have different influence on users, a user assigned with the top $5\%$ agent will be more likely to buy online insurance.
\end{itemize}

Given the above findings, we argue that designing a recommendations system specifically for online insurance is essential. It is also worth noting that, to give more accurate recommendations, we should try to represent the products accurately and take the influence of nonfinancial domain and agents into consideration. 

\begin{table}[!t]
\setlength{\abovecaptionskip}{0cm}
\setlength{\belowcaptionskip}{-0.2cm}
\centering
\caption{Notations and descriptions}
\label{notations}
\begin{tabular}{ll}
\toprule[1pt]
Notations       &     Descriptions\\
\hline
$\mathcal{D}^{s}$, $\mathcal{D}^{t}$                              & source domain and target domain \\
$\mathcal{U}$                                                                        & overlapping users in the two domains \\
$\boldsymbol{Y}^{s}$, $\boldsymbol{Y}^{t}$                  & rating matrices of source and target domain \\
$\mathcal{NI}_{u}^{s}$, $\mathcal{NI}_{u}^{t}$           & interacted item sequences of user $u$ in source\\
                                                                                                 & and target domain\\
$\mathcal{N}_{1}(e)$                                                           & $e$'s one-hop neighbors\\
$\mathcal{N}_{\rho}(e)$                                                       & the set of nodes connecting to $e$ by meta-path $\rho$\\
\bottomrule[1.0pt]
\end{tabular}
\end{table}

\subsection{Preliminary}
A heterogeneous information network (HIN) is a special kind of information network, which contains either multiple types of objects or multiple types of relations, which can be defined as follows:
\begin{definition}[Heterogeneous Information Network (HIN) \cite{DBLP:journals/pvldb/SunHYYW11}] 
A HIN is defined as a directed graph $\mathcal{G}=(\mathcal{V}, \mathcal{E})$ with an node type mapping function $\phi:\mathcal{V}\rightarrow\mathcal{A}$ and a relation type mapping function $\varphi: \mathcal{E}\rightarrow\mathcal{R}$. $\mathcal{A}$ and $\mathcal{R}$ denote the sets of predefined node and relation types, where $|\mathcal{A}|+|\mathcal{R}|>2$.
\end{definition}

In HINs, two objects can be connected via different semantic paths, which are called meta-paths.
\begin{definition}[Meta-path \cite{DBLP:journals/pvldb/SunHYYW11}] 
A meta-path $\rho$ is defined as a path in the form of $e_{1}\xrightarrow{r_{1}} e_{2}\xrightarrow{r_{2}}\ldots \xrightarrow{r_{L-1}}e_{L}$ (abbreviated as $e_{1},e_{2},\ldots, e_{l}$), which describes a composite relation $r = r_{1}\circ r_{2}\circ \ldots \circ r_{L-1}$ between object $e_{1}$ and $e_{L}$, where $\circ$ denotes the composition operator on relations.
\end{definition}

\section{Problem Formulation}
In this section, we formally define our problem, and summarize the notations and descriptions in Table \ref{notations}. As mentioned above, we have two domains, a source domain (nonfinancial domain) and a target domain (insurance domain). Let $\mathcal{U}=\{u_{1},u_{2},\ldots,u_{m}\}$ denote overlapping users between nonfinancial domain $\mathcal{D}^{s}$ and insurance domain $\mathcal{D}^{t}$, respectively. If a user only appears in one domain, he/she is a cold start user in the other domain. The user-item interaction matrices are denoted as $\boldsymbol{Y}^{s}\in\mathbb{R}^{m\times s}$ and $\boldsymbol{Y}^{t}\in\mathbb{R}^{m\times t}$, which are defined according to users’ implicit feedbacks. We additionally use $\mathcal{NI}_{u}^{s}$ and $\mathcal{NI}_{u}^{t}$ for the sequences of items that user $u$ has interacted with. 
Besides, the interactions in insurance domain can be abstracted as a heterogeneous information network (HIN), which we will illustrate later. Given rating matrices and HIN, our goal is to learn more effective latent features for users and items, and then learn the mapping function from nonfinancial domain to insurance domain, which can help us deal with cold start users.

\section {HCDIR}

\begin{figure}[!h]
\setlength{\abovecaptionskip}{0cm}
\setlength{\belowcaptionskip}{-0.4cm}
  \centering
  \includegraphics[scale=0.44]{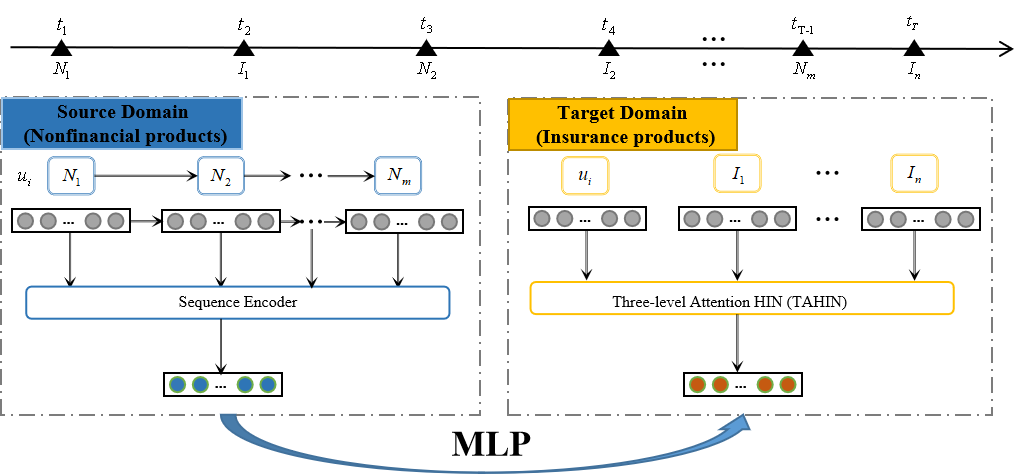}
  \caption{The Framework of HCDIR}
  \label{SIGIR2020_Industry_FRAMEWORK}
\end{figure}

To provide recommendations to cold start users, we propose HCDIR. As shown in Figure \ref{SIGIR2020_Industry_FRAMEWORK}, HCDIR contains three main parts: learning latent features of users in both insurance domain and nonfinancial domain, mapping of user latent features.

\subsection{Latent Feature in Insurance Domain}
\label{SIGIR2020_Industry_HIN}

\begin{figure*}[!t]
\setlength{\abovecaptionskip}{0.cm}
\setlength{\belowcaptionskip}{-0.4cm}
  \centering
  \includegraphics[scale=0.5]{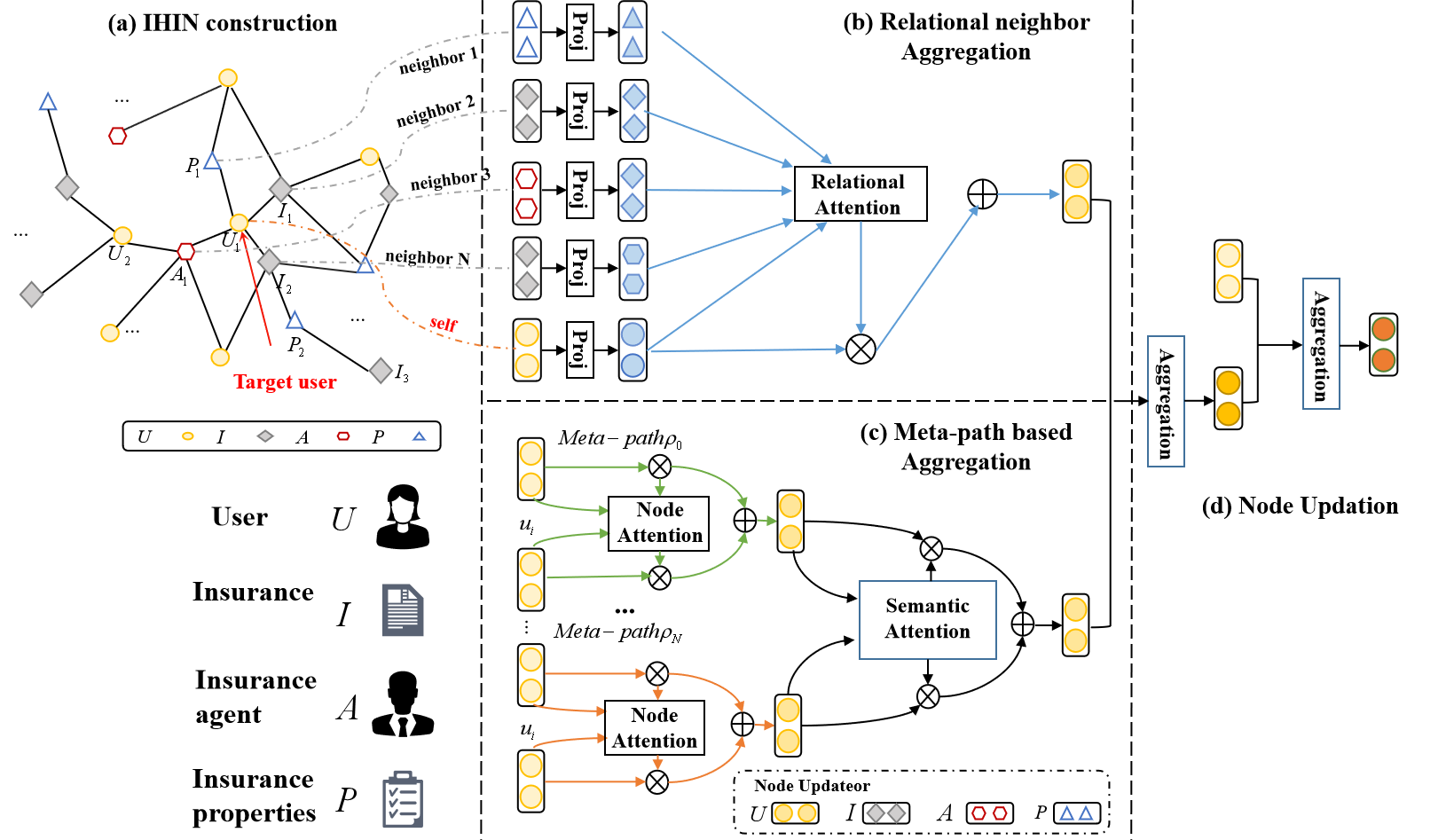}
  \caption{The Details of TAHIN (take node $U$ as example).  (a) illustrates the HIN constructed on Jinguanjia dataset. (b) is relational neighbors aggregation, we first project the neighbors to the same node type space, and aggregate them by calculating the weighted sum of one-hop neighbors. (c) is the node and semantic attention aggregation, the left part is to aggregate the meta-paths based neighbors, the right part is to aggregate the results from the left part. (d) is node updation, aggregating information from (b) and (c) to the original node representation.}
  \label{SIGIR2020_Industry_framework_HIN}
\end{figure*}

As mentioned above, the complexity of insurance products is typically non-trivial, understanding the items may require a considerable cognitive overload \cite{DBLP:conf/sac/RokachSSCS13}. Under this circumstance, generating efficient user embeddings is challenging. To achieve this goal, we design a three-level attention aggregation HIN method (TAHIN). In this part, we first introduce the IHIN we constructed based on Jinguanjia dataset, and then present how to learn effective user representations over the constructed IHIN. Figure \ref{SIGIR2020_Industry_framework_HIN} shows the details of TAHIN module. Specifically, we first propose relational attention to aggregate one-hop heterogeneous neighbors, and then node attention to aggregate meta-paths based neighbors, and semantic attention to aggregate meta-paths based neighbor sets. Finally, we aggregate the results of relational attention aggregation and semantic attention aggregation to the original node embedding to update node representations.

\subsubsection{Insurance Heterogeneous Information Network Construction} 
Interactions in insurance domain can be abstracted as an insurance heterogeneous information network (IHIN). Specifically, we define four types of nodes corresponding to user (U), agent (A), insurance product (I) and insurance product property (P), and six types of edges denoting various types of relations between them. As we mentioned above, insurance products are very complex, customers usually couldn't understand the whole insurance policies by just reading insurance products titles. Insurance products have several properties, which meet different demands for different customers. Therefore, we treat insurance property as a type of node. In this paper, we choose several properties customers most care about, which are price, level of assurance, character, coverage time, insurance type, age restriction, extra characters, etc. 
The schema of IHIN is displayed in Figure \ref{SIGIR2020_Industry_framework_HIN}(a), which is formally defined as:
\begin{definition}[Insurance Heterogeneous Information Network] 
\label{SIGIR2020_Industry_IHINdef}
Insurance Heterogeneous Information Network (IHIN) in our work is a HIN, containing four types of nodes: users $e_{u}$, agents $e_{a}$, insurance products $e_{i}$ and insurance product properties $e_{p}$. Edges exit between $e_{u}$ and $e_{a}$ denoting \emph{be served by} $r_{ua}$ and \emph{serve} $r_{au}$ relations, between $e_{u}$ and $e_{i}$ denoting \emph{purchase} $r_{ui}$ and \emph{be purchased by} $r_{iu}$ relations,
between $e_{i}$ and $e_{p}$ denoting \emph{possess} $r_{ip}$ and \emph{be possessed by} $r_{pi}$ relations.
\end{definition}

PingAn company possesses the data of user portrait, agent portrait and item portrait, for efficiency, we initialize IHIN using these data instead of initializing them randomly.

In IHIN, two nodes can be connected via different meta-paths. As shows in Figure  \ref{SIGIR2020_Industry_framework_HIN}, two insurance products can be connected via multiple meta-paths, e.g. insurance product-user-insurance product (I-U-I), insurance product-insurance property-insurance product (I-P-I), etc. Different meta-paths may reveal different semantics. For example, I-U-I means the two insurance products are needed by the same user, they may be complementary. I-P-I means the two insurance products have same properties, e.g. high level assurance. In addition, meta-paths can also connect different types of nodes, for example, user-insurance product-insurance property (U-I-P), which implies that the user bought the insurance, since she may concern most about the insurance property. Now, we can give the definition of meta-path based neighbors:
\begin{definition}[Meta-path based Neighbors \cite{DBLP:conf/aaai/HuZSZLQ19}]
Given a node $e$ and a meta-path $\rho$ in a HIN, the meta-path based neighbors $\mathcal{N}_{\rho}(e)$ of node $e$ is defined as a set of nodes which connect with node $e$ via meta-path $\rho$. Note that the node's meta-path based neighbors may have different node types.
\end{definition}

\subsubsection{Relational Neighbor Aggregation}
 
As different relations imply different information, as shown in Figure \ref{SIGIR2020_Industry_framework_HIN}(a), $I_{1}$ and $A_{1}$ are all neighbors of $U_{1}$, but they imply different information. We employ a relational attention aggregation over one-hop neighbors. Figure \ref{SIGIR2020_Industry_framework_HIN} (b) illustrates the framework. Specifically, instead of using the same aggregation function among different one-hop neighbors, we learn a specific aggregation function for each type of relation. Let $\boldsymbol{h}^{0}_{e}$ denote the current embedding of node $e$, as node's one-hop neighbors may have different node type with the node, so we first project them to the same node space ($\boldsymbol{P}_{r}$ is projection matrix), and then calculate the attention score:
\begin{eqnarray*}
\label{SIGIR2020_Industry_ratten}
\alpha_{ew}=\frac{\exp(f_{r}(\boldsymbol{h}_{e}^{0}, \boldsymbol{P}_{r}\boldsymbol{h}_{w}^{0}))}{\sum_{j\in\mathcal{N}_{1}(e)}\exp(f_{r}(\boldsymbol{h}_{e}^{0},\boldsymbol{P}_{r}\boldsymbol{h}_{j}^{0}))},\;\;\forall w\in\mathcal{N}_{1}(e),
\end{eqnarray*}
where $f_{r}(\cdot, \cdot)$ is the deep neutral network performing relational attention, $\alpha_{ew}$ is the level of influence of node $e_{w}$, $\mathcal{N}_{1}(e)$ is node $e$'s one-hop neighbors. Then, we aggregate information from $\mathcal{N}_{1}(e)$:
\begin{eqnarray}
\label{SIGIR2020_Industry_raggre}
\boldsymbol{h}_{e}^{1} =\sigma\left(\sum_{w\in\mathcal{N}_{1}(e)}\alpha_{ew}\boldsymbol{h}_{w}^{0}\right),
\end{eqnarray}
where $\sigma$ denotes the activation function.

\subsubsection{Meta-path based Aggregation} 
Two nodes can also be connected by meta-paths, since different meta-path based neighbors imply different information (e.g. information from insurance product $I_{2}$ to insurance product $I_{1}$ ($I_{2}$-$U_{1}$-$I_{1}$) is different from information from it to $I_{3}$ ($I_{2}$-$P_{2}$-$I_{3}$), as shown in figure \ref{SIGIR2020_Industry_framework_HIN}(a), since the former implies the same user, and the later implies the same property). For efficiency, we only choose the meta-path based neighbors that have the same node type with the node. The attention score is defined as:
\begin{eqnarray*}
\label{SIGIR2020_Industry_atten}
\beta^{\rho}_{ew}=\frac{\exp(f_{\rho}(\boldsymbol{h}_{e}^{0}, \boldsymbol{h}_{w}^{0}))}{\sum_{j\in\mathcal{N}_{\rho}(e)}\exp(f_{\rho}(\boldsymbol{h}_{e}^{0},\boldsymbol{h}_{j}^{0}))},\;\forall w\in\mathcal{N}_{\rho}(e),
\end{eqnarray*}
where $f_{\rho}(\cdot, \cdot)$ is the deep neutral network which performs the node-level attention,
$\beta_{ew}$ is the level of influence of node $e_{w}$. Then, we employ attention mechanism to aggregate the information of the meta-path based neighbors:
\begin{eqnarray*}
\label{SIGIR2020_Industry_aggre}
\boldsymbol{h}_{e}^{\rho} = \sigma\left(\sum_{w\in\mathcal{N}_{\rho}(e)}\beta^{\rho}_{ew}\boldsymbol{h}_{w}^{0}\right),
\end{eqnarray*} 
where $\sigma$ denotes the activation function. The procedure is illustrated in the left part of Figure \ref{SIGIR2020_Industry_framework_HIN}(c).

Given the meta-path set $\{\rho_{1},\rho_{2},\ldots,\rho_{N}\}$, after node attention aggregation, for node $e$, we can obtain $N$ node-level embeddings, denoted as $\{\boldsymbol{h}^{\rho_{1}}_{e}, \boldsymbol{h}^{\rho_{2}}_{e},\ldots,\boldsymbol{h}^{\rho_{N}}_{e}\}$. All the node embeddings are denoted as $\{E_{\rho_{1}}, E_{\rho_{2}},\ldots,E_{\rho_{N}}\}$. In the following part, we introduce how to aggregate these node-level embeddings.

To learn a more accurate node embedding, we try to fuse multiple node embeddings. Taking $\{E_{\rho_{1}}, E_{\rho_{2}},\ldots,E_{\rho_{N}}\}$ as input, as shown in the right part of Figure \ref{SIGIR2020_Industry_framework_HIN}(c), we first calculate the importance of each meta-path $\rho_{j}$:
\begin{eqnarray*}
\setlength{\abovedisplayskip}{-0.1pt}
\setlength{\belowdisplayskip}{0pt}
w_{\rho_{j}}=\frac{1}{|\mathcal{V}|}\sum_{e\in\mathcal{V}}\boldsymbol{q}^{T}\cdot\tanh(\boldsymbol{W}\boldsymbol{h}_{e}^{\rho_{j}}+b),
\end{eqnarray*}
and the weight for $\rho_{j},\;(j=1:N)$ is defined as:
\begin {eqnarray*}
\gamma_{\rho_{j}}=\frac{\text{exp}(w_{\rho_{j}})}{\sum_{j=1}^{N}\text{exp}(w_{\rho_{j}})}.
\end{eqnarray*}
Form the definition of the attention score, we can see that the higher $\gamma_{\rho_{j}}$, the more important meta-path $\rho_{j}$ is. Now, we can fuse these node-level embeddings to obtain the final node embeddings:
\begin{eqnarray}
\label{SIGIR2020_Industry_saggre}
\boldsymbol{h}^{2}_{e}=\sum_{j=1}^{N}\gamma_{\rho_{j}}\cdot\boldsymbol{h}^{\rho_{j}}_{e}.
\end{eqnarray}

\subsubsection{Node Updation}

Finally, we aggregate the information to node $e$ from $\boldsymbol{h}_{e}^{1}$ (from \eqref{SIGIR2020_Industry_raggre}) and $\boldsymbol{h}_{e}^{2}$ (from \eqref{SIGIR2020_Industry_saggre}):
\begin{eqnarray*}
\label{SIGIR2020_Industry_opt}
\boldsymbol{h}_{e}=\text{ReLU}\{\boldsymbol{W}_{2}\text{concat}[\boldsymbol{h}^{0},\boldsymbol{W}_{1}\text{concat}(\boldsymbol{h}_{e}^{1},\boldsymbol{h}_{e}^{2})+\boldsymbol{b}_{1}]+\boldsymbol{b}_{2}\}.
\end{eqnarray*}

After updating the HIN node embeddings, we can get the user and insurance product embedding, which are denoted as $\boldsymbol{u}^{t}$ and $\boldsymbol{v}^{t}$, respectively. The objective function in target domain is:
\begin{eqnarray}
\label{SIGIR2020_Industry_RS}
\mathcal{L}_{T}=\sum_{(u,v)\in Y^{t}}-\left(y_{uv}\log\hat{y}_{uv}+(1-y_{uv})\log(1-\hat{y}_{uv})\right),
\end{eqnarray}
where $
\hat{y}_{uv}=\sigma(f(\boldsymbol{u}^{t},\boldsymbol{v}^{t}))
$,
$\sigma(\cdot)$ is sigmoid function, $f$ is a ranking function which can be a dot-product or a deep neural network.

\subsection{Latent Feature in Nonfinancial Domain}
In Jinguanjia, each item $i$ in nonfinancial domain is associated with a description $c_{i}$. In order to learn more effective latent features, we employ word2vec \cite{DBLP:conf/nips/MikolovSCCD13}. Suppose there are $n$ words in $i$'s content $c_{i}$. Then we utilize word2vec to obtain word vectors, which are represented as $\{\boldsymbol{w}^{i}_{k}\}_{k=1}^{n}$. Then we concatenate word vectors and apply a max pooling over it to get the final item embedding:
\begin{eqnarray*}
\boldsymbol{i}=\text{max-pooling}(\text{concat}(\{\boldsymbol{w}^{i}_{k}\}_{k=1}^{n})).
\end{eqnarray*}

To model the final user latent feature $\boldsymbol{u}^{s}$, we employ GRU over the user's interacted sequence $\mathcal{NI}_{u}^{s}$,
\begin{eqnarray*}
\setlength{\abovedisplayskip}{0pt}
\setlength{\belowdisplayskip}{0pt}
\begin{aligned}
\label{SIGIR20_Industry_GRU}
\boldsymbol{x}_{n}=\;&\sigma(\boldsymbol{W}_{x}\boldsymbol{i}_{n}+\boldsymbol{U}_{x}\boldsymbol{h}_{n-1}+\boldsymbol{b}_{x})\\
\boldsymbol{r}_{n}=\;&\sigma(\boldsymbol{W}_{r}\boldsymbol{i}_{n}+\boldsymbol{U}_{r}\boldsymbol{h}_{n-1}+\boldsymbol{b}_{r})\\
\widetilde{\boldsymbol{h}}_{n} = \;&\tanh(\boldsymbol{W}_{h}\boldsymbol{i}_{n}+\boldsymbol{r}_{n}\circ\boldsymbol{U}_{h}\boldsymbol{h}_{n-1}+\boldsymbol{b}_{h})\\
\boldsymbol{h}_{n} = \;&(\boldsymbol{1}-\boldsymbol{x}_{n})\circ \boldsymbol{h}_{n-1}+\boldsymbol{x}_{n}\circ\widetilde{\boldsymbol{h}}_{n},
\end{aligned}
\end{eqnarray*}
where $\sigma$ is sigmoid function,
$\circ$ is element-wise product, $\boldsymbol{W}_{x}$, $\boldsymbol{W}_{r}$, $\boldsymbol{W}_{h} \in\mathbb{R}^{n_{H}\times d}$, $\boldsymbol{U}_{x}$, $\boldsymbol{U}_{r}$, $\boldsymbol{U}_{h}\in\mathbb{R}^{n_{H}\times n_{H}}$, $n_{H}=d$ is hidden size. And use $\boldsymbol{h}_{n}$ to represent the user, i.e. $\boldsymbol{u}^{s}=\boldsymbol{h}_{n}$. The loss function is the same as eq. \eqref{SIGIR2020_Industry_RS}, where $\hat{y}_{uv}=\sigma(f(\boldsymbol{u}^{s},\boldsymbol{v}^{s}))$.

\subsection{Mapping Function Between Two Domains}

Similar to study \cite{DBLP:conf/ijcai/ManSJC17}, we employ MLP to perform latent space matching from source domain to target domain. We take $\boldsymbol{u}^{s}$ as input and $\boldsymbol{u}^{t}$ as output. and the loss function can be formalized as:
\begin{eqnarray*}
\setlength{\abovedisplayskip}{2pt}
\setlength{\belowdisplayskip}{0pt}
\mathcal{L}_{\text{cross}} = \sum_{u\in\mathcal{U}}\|f_{\text{mlp}}(\boldsymbol{u}^{s})-\boldsymbol{u}^{t}\|_{2}.
\end{eqnarray*}

\subsection{Recommendation for Cold Start Users}

In this paper, we assume cold start users have interactions in nonfinancial domain, but no interactions in insurance domain. After learning the latent features in nonfinancial domain $\boldsymbol{u}^{s}$, we can get the corresponding mapping latent features $\hat{\boldsymbol{u}}^{t}=f_{\text{mlp}}(\boldsymbol{u}^{s})$. Based on learned $\hat{\boldsymbol{u}}^{t}$, we can make recommendations to cold start users.

\section{Experiment}

To evaluate the performance of HCDIR, we conduct extensive experiments and online A/B test on Jinguanjia dataset to answer the following key questions:

\textbf{RQ1}: How does our proposed HCDIR model perform compared with the state-of-the-art methods for CDR task?

\textbf{RQ2}: Can the proposed HCDIR alleviate the data sparsity problem in the target domain?

\textbf{RQ3}: How does different types of heterogeneous auxiliary information and other HIN options  affect the recommendation performance in HCDIR?

\subsection{Datasets}
As described in Section \ref{SIGIR2020_Industry_data_des}, we build and release a suitable dataset for insurance product recommendation task. We randomly split the overlapping users of Jinguanjia dataset into training set (60$\%$) to learn  parameters, validation set (20$\%$) to tune hyper-parameters, and testing set (20$\%$) for the final performance comparison. For the testing set, we remove their information in the target domain to utilize them as cold start users for evaluating the recommendation performance (i.e., test users). To study the performance changes of our proposed methods with respect to the number of overlapping users, we restrict the number of the overlapping users similarly to the real-world distribution. We build four training sets with a certain fraction $\eta\in\{10\%, 20\%, 50\%, 100\%\}$ of the overlapping users who do not belong to the test users in baseline comparison study.

\subsection{Baseline Models and Metrics}
Four widely used recommendation algorithms are compared with the variants of HCDIR. These baselines can be divided into two groups: (1) \textit{Single-domain Models}: BPR \cite{DBLP:journals/corr/abs-1205-2618} and GRU4REC \cite{DBLP:journals/corr/HidasiKBT15}; (2) \textit{Cross-domain Models}: EMCDR-BPR \cite{DBLP:conf/ijcai/ManSJC17}, EMCDR-GRU, two variants of HCDIR and HCDIR. The first group is utilized to validate the usefulness of cross-domain recommendation models, and the second group is used to demonstrate the advantage of TAHIN module in insurance domain to deal with various kinds of heterogeneous information including user purchase logs, agent and complex insurance products' properties. How to utilize different types of heterogeneous information is one of the key factors to boost the effectiveness of model. RGCN and HAN are two representative methods in handling heterogeneous data. Here, we designed two variants, HCDIR-RGCN and HCDIR-HAN adopting RGCN and HAN, respectively. HAN is superior to the other deep heterogeneous network embedding models such as Metapath2Vec. RGCN employs the relation-aggregators-based GCN to heterogeneous information network. HCDIR both leverages HAN and RGCN to process various kinds of heterogeneous information in insurance domian for better user representation, which can effectively improve the recommendation performance.

We evaluate all models with NDCG and Rec@N (N=1,3,5), which effectively evaluate the performance of recommendation methods. NDCG is used to observe the overall performance in terms of ranking insurances, while Recall@N is used to judge how accurately recommend insurances at top N positions.

\textbf{NDCG}: Normalized Discounted Cumulative Gain (NDCG) extends HR by assigning higher scores to the hits at higher positions in the ranking list.

\textbf{Recall@N(Rec@N)}: The primary evaluation metric is Recall, which measures the proportion of cases when the relevant item is amongst the top ranked items in all test cases.

\begin{table}[!t]
\setlength{\abovecaptionskip}{0.cm}
\setlength{\belowcaptionskip}{-0.5cm}
\centering
\footnotesize{
\caption{Performance comparison.}
\label{baseline}
\begin{tabular}{c|c|c|c|c|c|c}
\toprule[1pt]
\multicolumn{3}{c|}{Jinguanjia dataset}&\multicolumn{4}{c}{Metrics}\\
\hline
$\eta$	&	Group	&	Method	&	NDCG	&	Rec@1	&	Rec@3	&	Rec@5	\\
\hline
\multirow{7}{*}{10$\%$}
    &	Single-domain 	    &	BPR	&	0.0719	&	0.0213	&	0.0737	&	0.1248	\\
\cline{3-7}
	&	RS	&	GRU4REC	&	0.0036	&	0.0017	&	0.0032	&	0.0057	\\
\cline{2-7}
	&		&	EMCDR-BPR	&	0.0881	&	0.0324	&	0.0689	&	0.1543	\\
\cline{3-7}
	&	Cross-domain	&	EMCDR-GRU	&	0.1013	&	0.0284	&	0.0961	&	0.2088	\\
\cline{3-7}
	&	 RS	&	HCDIR-RGCN	&	0.2468	&	0.0967	&	0.3448	&	0.3849	\\
\cline{3-7}
	&		&	HCDIR-HAN	&	0.3206	&	0.1236	&	0.3476	&	0.4828	\\
\cline{3-7}
	&		&	HCDIR	&	0.3674	&	0.1366	&	0.4002	&	0.5543	\\
\hline
\multirow{7}{*}{20$\%$}
	&	Single-domain 	&	BPR	&	0.0789	&	0.0241	&	0.0864	&	0.1348	\\
\cline{3-7}
	&	RS	&	GRU4REC	&	0.0042	&	0.0022	&	0.0047	&	0.0061	\\
\cline{2-7}
	&		&	EMCDR-BPR	&	0.0984	&	0.0347	&	0.0848	&	0.1611	\\
\cline{3-7}
	&	Cross-domain	&	EMCDR-GRU	&	0.1112	&	0.0366	&	0.1308	&	0.2257	\\
\cline{3-7}
	&	 RS	&	HCDIR-RGCN	&	0.2579	&	0.1003	&	0.3516	&	0.4002	\\
\cline{3-7}
	&		&	HCDIR-HAN	&	0.3311	&	0.1273	&	0.3656	&	0.4927	\\
\cline{3-7}
	&		&	HCDIR	&	0.3769	&	0.1548	&	0.4189	&	0.5683	\\
\hline
\multirow{7}{*}{50$\%$}
    &	Single-domain	&	BPR	&	0.0791	&	0.0274	&	0.1205	&	0.1735	\\
\cline{3-7}
	&	 RS	&	GRU4REC	&	0.0117	&	0.0027	&	0.0114	&	0.0213	\\
\cline{2-7}
	&		&	EMCDR-BPR	&	0.1125	&	0.0402	&	0.1609	&	0.2281	\\
\cline{3-7}
	&	Cross-domain &	EMCDR-GRU	&	0.1289	&	0.0496	&	0.1594	&	0.2589	\\
\cline{3-7}
	&	RS	&	HCDIR-RGCN	&	0.2701	&	0.1166	&	0.3611	&	0.4341	\\
\cline{3-7}
	&		&	HCDIR-HAN	&	0.3432	&	0.1341	&	0.3946	&	0.5372	\\
\cline{3-7}
	&		&	HCDIR	&	0.3895	&	0.1636	&	0.4354	&	0.5827	\\
\hline
\multirow{7}{*}{100$\%$}
 	&	Single-domain 	&	BPR	&	0.1009	&	0.0354	&	0.1627	&	0.1809	\\
\cline{3-7}
	&	RS	&	GRU4REC	&	0.0137	&	0.0054	&	0.0154	&	0.0221	\\
\cline{2-7}
	&	    &	EMCDR-BPR	&	0.1359	&	0.0511	&	0.2059	&	0.2556	\\
\cline{3-7}
	&	Cross-domain 		&	EMCDR-GRU	&	0.1498	&	0.0806	&	0.2124	&	0.2486	\\
\cline{3-7}
	&	RS&	HCDIR-RGCN	&	0.3067	&	0.1247	&	0.3739	&	0.4974	\\
\cline{3-7}
	&		&	HCDIR-HAN	&	0.3703	&	0.1357	&	0.4254	&	0.5627	\\
\cline{3-7}
	&		&	HCDIR	&	0.4109	&	0.1873	&	0.4654	&	0.6128	\\
\bottomrule[1.0pt]
\end{tabular}
}
\end{table}

\subsection{Model Implement Details}
\textbf{Parameter Setting.} In TAHIN’s `relational neighbor aggregation’ part, message passing is set as mean operation and type wise reducer is set as sum operation. In TAHIN’s `meta-path based aggregation’ part, the meta-paths used here can be categorized into four groups according to node types. User meta-paths are [U I U], [U A U] and [U I P I U], item meta-paths are [I U I], [I P I] and [I U A U I], agent meta-paths are [A U A] and [A U I U A] and insurance products’ properties meta-paths are [P I P] and [P I U I P]. The number of attention head in GAT is set to 8. Owing to separate training in three tasks (insurance domain, nonfinancial domain and cross domain) in cold start scenario, single type of meta-paths cannot significantly affect the model performance while incorporation of all kinds of meta-paths can boost the performance. Final embedding dimension S is a key parameter in HCDIR discussed in the below section. We set the GRU hidden state size to 32 due to storage. We take Adam as our optimizing algorithm. For the hyper-parameters of the Adam optimizer,we set the learning rate $\alpha$= 0.001. These settings are chosen with grid search on the validation set. To speed up the training and converge quickly, we use batch size as 32. We test the model performance on the validation set for every epoch. We implement the proposed method based on Pytorch and DGL \cite{wang2019dgl}.  All experiments are performed in Nvidia Tesla V100.

\begin{figure}[!h]
\setlength{\abovecaptionskip}{-0.2cm}
\setlength{\belowcaptionskip}{-0.2cm}
  \centering
  \includegraphics[scale=0.4]{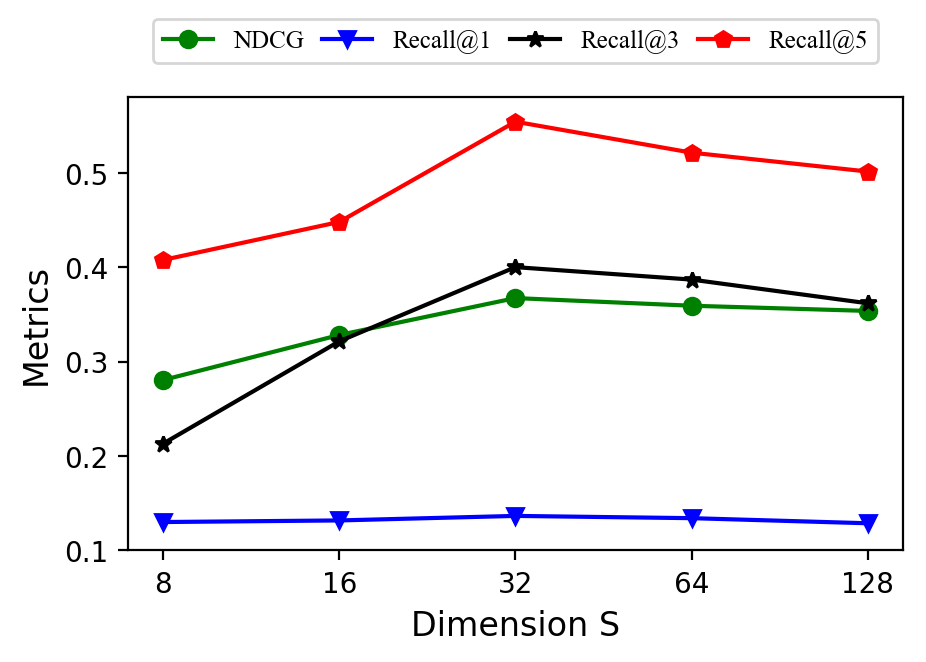}
  \caption{Study of the Final Embedding Dimension S.}
  \Description{Study of the Final Embedding Dimension S.}
  \label{SIGIR2020_Industry_S}
\end{figure}

\textbf{Study of the Final Embedding Dimension S.} The quality of the final emvedding can directly effect the performance of model. As shown in Figure \ref{SIGIR2020_Industry_S}, we can see that with the increase of the embedding dimension S, the performance raises first and then starts to drop slowly. The best parameter of S is 32. The main reason is that cross-domain method needs a suitable dimension to encode two domains' different information and larger dimension may introduce additional redundancies.

\subsection{Performance Comparison}
To answer RQ1 and RQ2, two variants of HCDIR are compared with four state-of-the-art models with different densities. Table \ref{baseline} shows the performance comparison. Overall, benefiting from the proposed TAHIN module and source domain information, HCDIR beats all comparative methods under all levels of data sparsity, respectively. These experiments reveal a number of interesting discoveries: (1) All cross-domain methods yield better performances than single-domain methods with mixture of target and source domain data , demonstrating the importance of cross-domain module; (2) Owing to the capability of using different types of heterogeneous information in insurance domian, two variants of HCDIR (HCDIR-RGCN and HCDIR-HAN) defeat other comparative methods; (3) HCDIR achieves a better performance in a sparser dataset compared with other methods. It is validated that, compared to comparative approaches, HCDIR can better alleviate the negative impacts of the data sparsity issue.

In order to anwser RQ3, we conduct experiments to compare HCDIR with HCDIR-RGCN and HCDIR-HAN. From the results of Table \ref{baseline}, we can find that the performance of HCDIR-RGCN and HCDIR-HAN declines sharply in terms of all the metrics when using a sparser dataset. This experiment shows that, the proposed HCDIR can get more stable and better performance with limited data, which mainly contributes various types of heterogeneous information and the incorporation of RGCN and HAN to deal with various kinds of auxiliary relationships.

\subsection{Ablation Study}
We find two important factors (data and corresponding data process module) effecting the performance of model. Therefore, we conduct the following two studies at 10$\%$ sparsity level, data ablation and model ablation study as shown in Table \ref{table2}.

\textbf{Result 1: Data Ablation}

In order to investigate the effect of two newly added heterogeneous data (Agents and Insurance Properties), we designed three variants of our proposed model, HCDIR with only interactions, HCDIR without Agent and HCDIR without Insurance Properties (short for HCDIR without IP). From Table \ref{table2},  it is found that only using interactions can not reach the best performance even use our proposed model framework. We can also observe that the performance of HCDIR without Agent declines more than HCDIR without IP compared to HCDIR in terms of all the metrics, which means agent heterogeneous information is the key factor to improve the model.

\textbf{Result 2: Model Ablation}

Two kinds of newly added heterogeneous information are used in HCDIR. How to leverage various types of heterogeneous information effectively may affect the final model performance. RGCN and HAN are two widely used methods in dealing with heterogeneous data, so we designed two variants of HCDIR in IHIN module, namely HCDIR using RGCN and HCDIR using HAN. HCDIR using HAN outperforms HCDIR using RGCN which aggregates 1-hop relation-aware neighbors. These results indicates the advantage of the combination attention mechanism and higher-order heterogeneous neighbors generated by GCN-based model in HAN. To improve HCDIR, we choose HAN and RGCN to deal with these heterogeneous information, which gains a better result.

\begin{table}[!t]
\centering
\footnotesize{
\caption{ Performance of variants of HCDIR on Jinguanjia dataset at 10$\%$ sparsity level}
\label{table2}
\begin{tabular}{c|c|c|c|c|c}
\toprule[1pt]
\multicolumn{2}{c|}{inguanjia dataset }&\multicolumn{4}{|c}{Metrics}\\
\cline{3-6}
\multicolumn{2}{c|}{at 10$\%$ sparsity level}&NDCG & Rec@1 &Rec@3 & Rec@5 \\
\hline

& $\#$HCDIR only	                & \multirow{1}[1]{*}{0.1013}    &\multirow{1}[1]{*}{0.0284}  	   &	\multirow{1}[1]{*}{0.0961} &	\multirow{1}[1]{*}{0.2088}\\
                                   &   with interactions  & \multirow{1}[2]{*}{\textbf{-72.43$\%$}}       & \multirow{1}[2]{*}{\textbf{-79.21$\%$}}  & \multirow{1}[2]{*}{\textbf{-75.99$\%$}}  & \multirow{1}[2]{*}{\textbf{-62.33$\%$}}\\
                                   \cline{2-6}
									Data   & $\#$HCDIR	                & \multirow{1}[1]{*}{0.2157}    &\multirow{1}[1]{*}{0.0933}  	   &	\multirow{1}[1]{*}{0.2287} &	\multirow{1}[1]{*}{0.3277}\\
                                Ablation   &  without Agent           & \multirow{1}[2]{*}{\textbf{-41.29$\%$}}       & \multirow{1}[2]{*}{\textbf{-31.70$\%$}}  & \multirow{1}[2]{*}{\textbf{-42.85$\%$}}  & \multirow{1}[2]{*}{\textbf{-40.88$\%$}}\\
										\cline{2-6}
                                   & $\#$HCDIR	                & \multirow{1}[1]{*}{0.2313}    &\multirow{1}[1]{*}{0.1073}  	   &	\multirow{1}[1]{*}{0.2512} &	\multirow{1}[1]{*}{0.3665}\\
                                   &  without IP  & \multirow{1}[2]{*}{\textbf{-37.04$\%$} }      & \multirow{1}[2]{*}{\textbf{-21.45$\%$}}  & \multirow{1}[2]{*}{\textbf{-37.23$\%$}}  & \multirow{1}[2]{*}{\textbf{-33.88$\%$}}\\

\hline 
 & $\#$HCDIR	                & \multirow{1}[1]{*}{0.2468}    &\multirow{1}[1]{*}{0.0967}  	   &	\multirow{1}[1]{*}{0.3448} &	\multirow{1}[1]{*}{0.3849}\\
                                 Model  &  using RGCN           & \multirow{1}[2]{*}{\textbf{-32.83$\%$}}       & \multirow{1}[2]{*}{\textbf{-29.21$\%$}}  & \multirow{1}[2]{*}{\textbf{-13.84$\%$}}  & \multirow{1}[2]{*}{\textbf{-30.56$\%$}}\\
										\cline{2-6}
                                Ablation   & $\#$HCDIR	                & \multirow{1}[1]{*}{0.3206}    &\multirow{1}[1]{*}{0.1236}  	   &	\multirow{1}[1]{*}{0.3476} &	\multirow{1}[1]{*}{0.4828}\\
                                   &  using HAN          & \multirow{1}[2]{*}{\textbf{-12.74$\%$}}       & \multirow{1}[2]{*}{\textbf{-9.52$\%$}}  & \multirow{1}[2]{*}{\textbf{-13.14$\%$}}  & \multirow{1}[2]{*}{\textbf{-12.90$\%$}}\\
\hline
\multirow{1}{*}{Full Model}                &	$\#$HCDIR	                           &	    0.3674        &	0.1366	         &	0.4002  &	0.5543\\
\bottomrule[1.0pt]
\end{tabular}
}
\end{table}
\subsection{Online A/B Testing for Cold-Start Recommendation}

\begin{table*}[!htbp]
\centering
%\small{
\caption{Online performance of compared methods. ‘G$\_$Baseline’ indicates the baseline performance of cold start user group using traditional method LightGBM; and ‘G$\_$HCDIR without agent’ and  ‘G$\_$HCDIR’denotes HCDIR without agent heterogeneous relationships and HCDIR, respectively.}
\label{table3}
\begin{tabular}{c|c|c|c|c|c|c}
\toprule[1pt]
\multirow{2}{*}{Metrics}&	\multicolumn{3}{|c|}{G$\_$HCDIR without agent}	&\multicolumn{3}{|c}{G$\_$HCDIR}\\	
\cline{2-7}	
     &	T+1 month	&	T+2 months	&	T+3 months	&	T+1 month	&	T+2 months	&	T+3 months	\\
\hline
improvement percentage 	& \multirow{2}[2]{*}{8.79$\%$} & \multirow{2}[2]{*}{12.87$\%$} & \multirow{2}[2]{*}{18.38$\%$} & \multirow{2}[2]{*}{12.94$\%$} & \multirow{2}[2]{*}{18.66$\%$}  & \multirow{2}[2]{*}{23.20$\%$}  \\
of UPCR vs G$\_$Baseline 	&       &       &       &       &       &  \\
\hline
improvement percentage 	& \multirow{2}[2]{*}{10.97$\%$} & \multirow{2}[2]{*}{13.04$\%$} & \multirow{2}[2]{*}{15.31$\%$} & \multirow{2}[2]{*}{15.25$\%$} & \multirow{2}[2]{*}{20.41$\%$}  & \multirow{2}[2]{*}{25.62$\%$}  \\
of UPGR vs G$\_$Baseline 	&       &       &       &       &       &  \\
\hline
 improvement percentage &	\multicolumn{3}{c|}{\multirow{2}*{ -79.94$\%$}}&\multicolumn{3}{|c}{\multirow{2}*{  -76.59$\%$}}\\
 of runing time vs G$\_$Baseline&  \multicolumn{3}{c|}{\multirow{2}*{  }}&\multicolumn{3}{|c}{\multirow{2}*{   }}\\
\bottomrule[1.0pt]
\end{tabular}
%}
\end{table*}

To validate the effectiveness of HCDIR, we implement online A/B test for insurance domain’s cold-start users to show how cross domain method and heterogeneous insurance information affect cold start recommendation in real-world scenario. 

For online A/B testing, cold-start users who haven’t purchased any insurance products by the end of August 2019 are divided into three groups with highly similar activities in Jinguangjia APP where each group contains 150,000 users. Users of the first group users are recommended insurance products by traditional strategy using best trained machine learning model LightGBM, donated as G$\_$Baseline. Users of the second group are recommended by HCDIR without agent heterogeneous information with 10$\%$ training data of Jinguanjia dataset used above, donated as G$\_$HCDIR without agent. Users of the third group are by our proposed HCDIR trained with 10$\%$ training data, donated as G$\_$HCDIR.

\textit{User Purchase Conversion Rate}(UPCR): Number of users who purchased the recommended insurance product divide total number of cold start users

\textit{User Premium Growth Amount} (UPGA): Amount of insurance premium cold start users paid for the recommended insurance.
 
Table \ref{table3} shows the results of our designed online A/B testing compared with G$\_$Baseline as baseline. We compare the performance of G$\_$HCDIR without agent to machine learning method G$\_$Baseline using LightGBM with only user-item interactions and designed features. From Table \ref{table3}, we find that the performance of G$\_$HCDIR without agent and G$\_$HCDIR using all the heterogeneous information consistently outperform these baseline methods. The improvement of UPCR and UPGR gradually increase over time, which indicates it need time for cold start users to develop insurance awareness. Specifically, it can be observed that G$\_$HCDIR without agent at least improves UPCR and UPGA by 8.79 $\%$ and 10.97 $\%$ in the time period from 1 month to 3 months compared to the traditional baseline G$\_$Baseline, respectively, which fully demonstrates the comprehensive effectiveness of TAHIN module in insurance domain and cross domain recommendation method. Moreover, with the help of ’agent’ heterogeneous auxiliary information, the improvements of UPCR and UPGR in G$\_$HCDIR are larger than that of G$\_$HCDIR without-agent. As mentioned above, agents are the key way to improve UPCR and UPGR in traditional insurance recommendation, and it also proves that the strong power of agent can significantly boost the performance of cold start problem in online insurance recommendation. 
As for training time, the training time of G$\_$Baseline model is 39.15 minutes, while the training time of G$\_$HCDIR without agent and G$\_$HCDIR are 7.86 minutes and 9.17 minutes, respectively. As shown in Table 6, our proposed HCDIR can at least improve by 76 $\%$.

\section{Related Work}
\subsection{Insurance Recommendation System}
To our knowledge, there are not many papers about recommendation systems in insurance products domain, some includes \cite{DBLP:conf/sac/RokachSSCS13,baobaohenshengqi2,baobaohenshengqi14,DBLP:conf/recsys/QaziFMF17,DBLP:conf/icmcs/LiuZKZZ019}. \cite{DBLP:conf/sac/RokachSSCS13} throughly describes the differences between recommendation system for classical domain and insurance domain, and focuses on call centers servicing Life and Annual insurance, where the agents also have limited knowledge and experience. \cite{baobaohenshengqi2} propose a web recommendation system for life insurance sector by using association rules, which is one of the most well researched techniques of data mining. \cite{baobaohenshengqi14} presents a hybird recommendation system in insurance domain based on a standard user-user collaborate filtering approach. \cite{DBLP:conf/recsys/QaziFMF17} utilizes Bayes networks to give customers personalized recommendation based on what other similar people with similar portfolios have.
\cite{DBLP:conf/icmla/KanchinadamQBMM18} is a improved model of \cite{DBLP:conf/recsys/QaziFMF17}, which tries to learn the structure of Bayesian network and considerably speeds up both training and inference run-times, while achieving similar accuracy.
\cite{DBLP:conf/icmcs/LiuZKZZ019} propose a causation-driven visualization system that fundamentally transforms cross-media insurance data into network diagrams and performs recommendation reasoning.
However, these methods neglect the item complexity and data sparsity problem.
\subsection{Cross-domain Recommendation}
Cross-domain recommendation (CDR) \cite{DBLP:conf/sigir/Wang0NC17,DBLP:conf/ijcai/ManSJC17,DBLP:conf/sigir/MaRLCMR19,DBLP:conf/cikm/KangHLY19,DBLP:conf/aaai/FuPWXL19,DBLP:conf/sigir/LinGL19}, which aims to improve the recommendation performance by means of transferring information from the auxiliary domain to the target domain, is one of the promising ways to solve data sparsity and cold start problem. Generally, CDR can be categorized into two categories. One is to aggregate knowledge between two domains, this kind of methods are interested in improving the overall performance in the target domain \cite{DBLP:conf/sigir/Wang0NC17,DBLP:conf/sigir/LinGL19,DBLP:conf/sigir/MaRLCMR19}, however, they can not deal with cold start users. Since cold start users do not have any interactions in target domain. The other one aims at infering the preferences of cold start users based on their preferences observed in other domains \cite{DBLP:conf/ijcai/ManSJC17,DBLP:conf/cikm/KangHLY19,DBLP:conf/aaai/FuPWXL19}. These methods assume that there exists overlap in information between users and/or items across different domains, and train a mapping function from the source-domain into the target-domain. For cold start users, these method first learn representations in source domain, and then mapping them to the target domain.

\subsection{Heterogeneous Information Networks}
Recently, some methods have been proposed representation learning methods for HIN. These methods can be grossly divided into two groups: shallow models and deep models. Shallow models (\cite{DBLP:conf/kdd/DongCS17,DBLP:conf/cikm/FuLL17,DBLP:conf/kdd/HuSZY18,DBLP:conf/aaai/LuSH019} employ factorization-­based approaches or random walk approaches to aggregate information from neighbor nodes. For example, Metapath2vec \cite{DBLP:conf/kdd/DongCS17} formalizes meta-paths based random walks to obtain heterogeneous neighborhoods of a node and leverages Skip-gram model to learn the network structure. However, this kind of method only explore one aspect information, failing to integrate more heterogeneous information. Deep models \cite{DBLP:conf/kdd/ZhangSHSC19,DBLP:conf/www/WangJSWYCY19} aggregate neighbor information by neural network based method. HetGNN \cite{DBLP:conf/kdd/ZhangSHSC19} jointly learn heterogeneous graph information and heterogeneous contents information for node embeddings based on GNN \cite{4700287}. Inspired by graph attention networks, R-GCNs \cite{DBLP:conf/esws/SchlichtkrullKB18} are developed to deal with the highly multi-relational data. HAN \cite{DBLP:conf/www/WangJSWYCY19} designs a two level (node-level and semantic-level) attentions to generate node embedding by aggregating features from meta-path based neighbors.

\section{Conclusion and Future Work}
To deal with insurance product complexity and cold start problem, we propose a novel framework called a HCDIR for cold start users in insurance domain. Specifically, we first try to learn more effective user and item latent features in both source and target domains. In source domain, we employ GRU to module users' dynamic interests. In target domain, we construct an IHIN based on data from Jinguanjia App, then we employ three-level (relational, node, and semantic) attention aggregations to get user and insurance product representations. After obtaining the latent features of the overlapping users, a feature mapping between the two domains is learned by MLP. We apply HCDIR on PingAn Jinguanjia dataset, and show HCDIR significantly outperforms the state-of-the-art solutions. As future work, we will try to construct more complete HIN, considering more types of relations, such as the relation between agent and insurance product. We will also consider to train more accurate item representations in source domain.

\bibliographystyle{ACM-Reference-Format}
\balance
\bibliography{Sigir_main}


\begin{thebibliography}{31}

%%% the Web address.

\ifx \showCODEN    \undefined \def \showCODEN     #1{\unskip}     \fi
\ifx \showDOI      \undefined \def \showDOI       #1{#1}\fi
\ifx \showISBNx    \undefined \def \showISBNx     #1{\unskip}     \fi
\ifx \showISBNxiii \undefined \def \showISBNxiii  #1{\unskip}     \fi
\ifx \showISSN     \undefined \def \showISSN      #1{\unskip}     \fi
\ifx \showLCCN     \undefined \def \showLCCN      #1{\unskip}     \fi
\ifx \shownote     \undefined \def \shownote      #1{#1}          \fi
\ifx \showarticletitle \undefined \def \showarticletitle #1{#1}   \fi
\ifx \showURL      \undefined \def \showURL       {\relax}        \fi
% invisible to TeX
\providecommand\bibfield[2]{#2}
\providecommand\bibinfo[2]{#2}
\providecommand\natexlab[1]{#1}
\providecommand\showeprint[2][]{arXiv:#2}

\bibitem[\protect\citeauthoryear{Abdollahi and Nasraoui}{Abdollahi and
  Nasraoui}{2016}]%
        {DBLP:conf/www/AbdollahiN16}
\bibfield{author}{\bibinfo{person}{Behnoush Abdollahi} {and}
  \bibinfo{person}{Olfa Nasraoui}.} \bibinfo{year}{2016}\natexlab{}.
\newblock \showarticletitle{Explainable Matrix Factorization for Collaborative
  Filtering}. In \bibinfo{booktitle}{\emph{Proceedings of the 25th
  International Conference on World Wide Web, {WWW} 2016, Montreal, Canada,
  April 11-15, 2016, Companion Volume}}. \bibinfo{pages}{5--6}.
\newblock
\urldef\tempurl%
\url{https://doi.org/10.1145/2872518.2889405}
\showDOI{\tempurl}


\bibitem[\protect\citeauthoryear{Chung, G{\"{u}}l{\c{c}}ehre, Cho, and
  Bengio}{Chung et~al\mbox{.}}{2014}]%
        {DBLP:journals/corr/ChungGCB14}
\bibfield{author}{\bibinfo{person}{Junyoung Chung},
  \bibinfo{person}{{\c{C}}aglar G{\"{u}}l{\c{c}}ehre},
  \bibinfo{person}{KyungHyun Cho}, {and} \bibinfo{person}{Yoshua Bengio}.}
  \bibinfo{year}{2014}\natexlab{}.
\newblock \showarticletitle{Empirical Evaluation of Gated Recurrent Neural
  Networks on Sequence Modeling}.
\newblock \bibinfo{journal}{\emph{CoRR}}  \bibinfo{volume}{abs/1412.3555}
  (\bibinfo{year}{2014}).
\newblock
\showeprint[arxiv]{1412.3555}
\urldef\tempurl%
\url{http://arxiv.org/abs/1412.3555}
\showURL{%
\tempurl}


\bibitem[\protect\citeauthoryear{Dong, Chawla, and Swami}{Dong
  et~al\mbox{.}}{2017}]%
        {DBLP:conf/kdd/DongCS17}
\bibfield{author}{\bibinfo{person}{Yuxiao Dong}, \bibinfo{person}{Nitesh~V.
  Chawla}, {and} \bibinfo{person}{Ananthram Swami}.}
  \bibinfo{year}{2017}\natexlab{}.
\newblock \showarticletitle{metapath2vec: Scalable Representation Learning for
  Heterogeneous Networks}. In \bibinfo{booktitle}{\emph{Proceedings of the 23rd
  {ACM} {SIGKDD} International Conference on Knowledge Discovery and Data
  Mining, Halifax, NS, Canada, August 13 - 17, 2017}}.
  \bibinfo{pages}{135--144}.
\newblock
\urldef\tempurl%
\url{https://doi.org/10.1145/3097983.3098036}
\showURL{%
\tempurl}


\bibitem[\protect\citeauthoryear{Fu, Lee, and Lei}{Fu et~al\mbox{.}}{2017}]%
        {DBLP:conf/cikm/FuLL17}
\bibfield{author}{\bibinfo{person}{Tao{-}Yang Fu},
  \bibinfo{person}{Wang{-}Chien Lee}, {and} \bibinfo{person}{Zhen Lei}.}
  \bibinfo{year}{2017}\natexlab{}.
\newblock \showarticletitle{HIN2Vec: Explore Meta-paths in Heterogeneous
  Information Networks for Representation Learning}. In
  \bibinfo{booktitle}{\emph{Proceedings of the 2017 {ACM} on Conference on
  Information and Knowledge Management, {CIKM} 2017, Singapore, November 06 -
  10, 2017}}. \bibinfo{pages}{1797--1806}.
\newblock
\urldef\tempurl%
\url{https://doi.org/10.1145/3132847.3132953}
\showDOI{\tempurl}


\bibitem[\protect\citeauthoryear{Fu, Peng, Wang, Xu, and Li}{Fu
  et~al\mbox{.}}{2019}]%
        {DBLP:conf/aaai/FuPWXL19}
\bibfield{author}{\bibinfo{person}{Wenjing Fu}, \bibinfo{person}{Zhaohui Peng},
  \bibinfo{person}{Senzhang Wang}, \bibinfo{person}{Yang Xu}, {and}
  \bibinfo{person}{Jin Li}.} \bibinfo{year}{2019}\natexlab{}.
\newblock \showarticletitle{Deeply Fusing Reviews and Contents for Cold Start
  Users in Cross-Domain Recommendation Systems}. In
  \bibinfo{booktitle}{\emph{The Thirty-Third {AAAI} Conference on Artificial
  Intelligence, {AAAI} 2019, Honolulu, Hawaii, USA, January 27 - February 1,
  2019}}. \bibinfo{pages}{94--101}.
\newblock
\urldef\tempurl%
\url{https://doi.org/10.1609/aaai.v33i01.330194}
\showDOI{\tempurl}


\bibitem[\protect\citeauthoryear{Gupta and Jain}{Gupta and Jain}{2013}]%
        {baobaohenshengqi2}
\bibfield{author}{\bibinfo{person}{Abdhesh Gupta} {and} \bibinfo{person}{Anwiti
  Jain}.} \bibinfo{year}{2013}\natexlab{}.
\newblock \showarticletitle{Life insurance recommender system based on
  association rule mining and dual clustering method for solving cold-start
  problem}.
\newblock \bibinfo{journal}{\emph{International Journal of Advanced Research in
  Computer Science and Software Engineering}}  \bibinfo{volume}{3}
  (\bibinfo{date}{Oct.} \bibinfo{year}{2013}).
\newblock


\bibitem[\protect\citeauthoryear{Hamilton, Ying, and Leskovec}{Hamilton
  et~al\mbox{.}}{2017}]%
        {DBLP:conf/nips/HamiltonYL17}
\bibfield{author}{\bibinfo{person}{William~L. Hamilton},
  \bibinfo{person}{Zhitao Ying}, {and} \bibinfo{person}{Jure Leskovec}.}
  \bibinfo{year}{2017}\natexlab{}.
\newblock \showarticletitle{Inductive Representation Learning on Large Graphs}.
  In \bibinfo{booktitle}{\emph{Advances in Neural Information Processing
  Systems 30: Annual Conference on Neural Information Processing Systems 2017,
  4-9 December 2017, Long Beach, CA, {USA}}}. \bibinfo{pages}{1024--1034}.
\newblock


\bibitem[\protect\citeauthoryear{Hidasi, Karatzoglou, Baltrunas, and
  Tikk}{Hidasi et~al\mbox{.}}{2016}]%
        {DBLP:journals/corr/HidasiKBT15}
\bibfield{author}{\bibinfo{person}{Bal{\'{a}}zs Hidasi},
  \bibinfo{person}{Alexandros Karatzoglou}, \bibinfo{person}{Linas Baltrunas},
  {and} \bibinfo{person}{Domonkos Tikk}.} \bibinfo{year}{2016}\natexlab{}.
\newblock \showarticletitle{Session-based Recommendations with Recurrent Neural
  Networks}. In \bibinfo{booktitle}{\emph{4th International Conference on
  Learning Representations, {ICLR} 2016, San Juan, Puerto Rico, May 2-4, 2016,
  Conference Track Proceedings}}.
\newblock
\urldef\tempurl%
\url{http://arxiv.org/abs/1511.06939}
\showURL{%
\tempurl}


\bibitem[\protect\citeauthoryear{Hu, Shi, Zhao, and Yu}{Hu
  et~al\mbox{.}}{2018}]%
        {DBLP:conf/kdd/HuSZY18}
\bibfield{author}{\bibinfo{person}{Binbin Hu}, \bibinfo{person}{Chuan Shi},
  \bibinfo{person}{Wayne~Xin Zhao}, {and} \bibinfo{person}{Philip~S. Yu}.}
  \bibinfo{year}{2018}\natexlab{}.
\newblock \showarticletitle{Leveraging Meta-path based Context for Top- {N}
  Recommendation with {A} Neural Co-Attention Model}. In
  \bibinfo{booktitle}{\emph{Proceedings of the 24th {ACM} {SIGKDD}
  International Conference on Knowledge Discovery {\&} Data Mining, {KDD} 2018,
  London, UK, August 19-23, 2018}}. \bibinfo{pages}{1531--1540}.
\newblock
\urldef\tempurl%
\url{https://doi.org/10.1145/3219819.3219965}
\showDOI{\tempurl}


\bibitem[\protect\citeauthoryear{Hu, Zhang, Shi, Zhou, Li, and Qi}{Hu
  et~al\mbox{.}}{2019}]%
        {DBLP:conf/aaai/HuZSZLQ19}
\bibfield{author}{\bibinfo{person}{Binbin Hu}, \bibinfo{person}{Zhiqiang
  Zhang}, \bibinfo{person}{Chuan Shi}, \bibinfo{person}{Jun Zhou},
  \bibinfo{person}{Xiaolong Li}, {and} \bibinfo{person}{Yuan Qi}.}
  \bibinfo{year}{2019}\natexlab{}.
\newblock \showarticletitle{Cash-Out User Detection Based on Attributed
  Heterogeneous Information Network with a Hierarchical Attention Mechanism}.
  In \bibinfo{booktitle}{\emph{The Thirty-Third {AAAI} Conference on Artificial
  Intelligence, {AAAI} 2019, Honolulu, Hawaii, USA, January 27 - February 1,
  2019}}. \bibinfo{pages}{946--953}.
\newblock
\urldef\tempurl%
\url{https://doi.org/10.1609/aaai.v33i01.3301946}
\showDOI{\tempurl}


\bibitem[\protect\citeauthoryear{Kanchinadam, Qazi, Bockhorst, Morell,
  Meissner, and Fung}{Kanchinadam et~al\mbox{.}}{2018}]%
        {DBLP:conf/icmla/KanchinadamQBMM18}
\bibfield{author}{\bibinfo{person}{Teja Kanchinadam}, \bibinfo{person}{Maleeha
  Qazi}, \bibinfo{person}{Joseph Bockhorst}, \bibinfo{person}{Mary~Y. Morell},
  \bibinfo{person}{Katie~J. Meissner}, {and} \bibinfo{person}{Glenn Fung}.}
  \bibinfo{year}{2018}\natexlab{}.
\newblock \showarticletitle{Using Discriminative Graphical Models for Insurance
  Recommender Systems}. In \bibinfo{booktitle}{\emph{17th {IEEE} International
  Conference on Machine Learning and Applications, {ICMLA} 2018, Orlando, FL,
  USA, December 17-20, 2018}}. \bibinfo{pages}{421--428}.
\newblock
\urldef\tempurl%
\url{https://doi.org/10.1109/ICMLA.2018.00069}
\showDOI{\tempurl}


\bibitem[\protect\citeauthoryear{Kang, Hwang, Lee, and Yu}{Kang
  et~al\mbox{.}}{2019}]%
        {DBLP:conf/cikm/KangHLY19}
\bibfield{author}{\bibinfo{person}{SeongKu Kang}, \bibinfo{person}{Junyoung
  Hwang}, \bibinfo{person}{Dongha Lee}, {and} \bibinfo{person}{Hwanjo Yu}.}
  \bibinfo{year}{2019}\natexlab{}.
\newblock \showarticletitle{Semi-Supervised Learning for Cross-Domain
  Recommendation to Cold-Start Users}. In \bibinfo{booktitle}{\emph{Proceedings
  of the 28th {ACM} International Conference on Information and Knowledge
  Management, {CIKM} 2019, Beijing, China, November 3-7, 2019}}.
  \bibinfo{pages}{1563--1572}.
\newblock
\urldef\tempurl%
\url{https://doi.org/10.1145/3357384.3357914}
\showDOI{\tempurl}


\bibitem[\protect\citeauthoryear{Lin, Gao, and Li}{Lin et~al\mbox{.}}{2019}]%
        {DBLP:conf/sigir/LinGL19}
\bibfield{author}{\bibinfo{person}{Tzu{-}Heng Lin}, \bibinfo{person}{Chen Gao},
  {and} \bibinfo{person}{Yong Li}.} \bibinfo{year}{2019}\natexlab{}.
\newblock \showarticletitle{{CROSS:} Cross-platform Recommendation for Social
  E-Commerce}. In \bibinfo{booktitle}{\emph{Proceedings of the 42nd
  International {ACM} {SIGIR} Conference on Research and Development in
  Information Retrieval, {SIGIR} 2019, Paris, France, July 21-25, 2019}}.
  \bibinfo{pages}{515--524}.
\newblock
\urldef\tempurl%
\url{https://doi.org/10.1145/3331184.3331191}
\showDOI{\tempurl}


\bibitem[\protect\citeauthoryear{Liu, Zang, Kuang, Zou, Zheng, and Cui}{Liu
  et~al\mbox{.}}{2019}]%
        {DBLP:conf/icmcs/LiuZKZZ019}
\bibfield{author}{\bibinfo{person}{Zhixiu Liu}, \bibinfo{person}{Chengxi Zang},
  \bibinfo{person}{Kun Kuang}, \bibinfo{person}{Hao Zou}, \bibinfo{person}{Hu
  Zheng}, {and} \bibinfo{person}{Peng Cui}.} \bibinfo{year}{2019}\natexlab{}.
\newblock \showarticletitle{Causation-Driven Visualizations for Insurance
  Recommendation}. In \bibinfo{booktitle}{\emph{{IEEE} International Conference
  on Multimedia {\&} Expo Workshops, {ICME} Workshops 2019, Shanghai, China,
  July 8-12, 2019}}. \bibinfo{pages}{471--476}.
\newblock
\urldef\tempurl%
\url{https://doi.org/10.1109/ICMEW.2019.00087}
\showDOI{\tempurl}


\bibitem[\protect\citeauthoryear{Lu, Shi, Hu, and Liu}{Lu
  et~al\mbox{.}}{2019}]%
        {DBLP:conf/aaai/LuSH019}
\bibfield{author}{\bibinfo{person}{Yuanfu Lu}, \bibinfo{person}{Chuan Shi},
  \bibinfo{person}{Linmei Hu}, {and} \bibinfo{person}{Zhiyuan Liu}.}
  \bibinfo{year}{2019}\natexlab{}.
\newblock \showarticletitle{Relation Structure-Aware Heterogeneous Information
  Network Embedding}. In \bibinfo{booktitle}{\emph{The Thirty-Third {AAAI}
  Conference on Artificial Intelligence, {AAAI} 2019, Honolulu, Hawaii, USA,
  January 27 - February 1, 2019}}. \bibinfo{pages}{4456--4463}.
\newblock
\urldef\tempurl%
\url{https://doi.org/10.1609/aaai.v33i01.33014456}
\showDOI{\tempurl}


\bibitem[\protect\citeauthoryear{Ma, Ren, Lin, Chen, Ma, and de~Rijke}{Ma
  et~al\mbox{.}}{2019}]%
        {DBLP:conf/sigir/MaRLCMR19}
\bibfield{author}{\bibinfo{person}{Muyang Ma}, \bibinfo{person}{Pengjie Ren},
  \bibinfo{person}{Yujie Lin}, \bibinfo{person}{Zhumin Chen},
  \bibinfo{person}{Jun Ma}, {and} \bibinfo{person}{Maarten de Rijke}.}
  \bibinfo{year}{2019}\natexlab{}.
\newblock \showarticletitle{{\(\pi\)}-Net: {A} Parallel Information-sharing
  Network for Shared-account Cross-domain Sequential Recommendations}. In
  \bibinfo{booktitle}{\emph{Proceedings of the 42nd International {ACM} {SIGIR}
  Conference on Research and Development in Information Retrieval, {SIGIR}
  2019, Paris, France, July 21-25, 2019}}. \bibinfo{pages}{685--694}.
\newblock
\urldef\tempurl%
\url{https://doi.org/10.1145/3331184.3331200}
\showDOI{\tempurl}


\bibitem[\protect\citeauthoryear{Man, Shen, Jin, and Cheng}{Man
  et~al\mbox{.}}{2017}]%
        {DBLP:conf/ijcai/ManSJC17}
\bibfield{author}{\bibinfo{person}{Tong Man}, \bibinfo{person}{Huawei Shen},
  \bibinfo{person}{Xiaolong Jin}, {and} \bibinfo{person}{Xueqi Cheng}.}
  \bibinfo{year}{2017}\natexlab{}.
\newblock \showarticletitle{Cross-Domain Recommendation: An Embedding and
  Mapping Approach}. In \bibinfo{booktitle}{\emph{Proceedings of the
  Twenty-Sixth International Joint Conference on Artificial Intelligence,
  {IJCAI} 2017, Melbourne, Australia, August 19-25, 2017}}.
  \bibinfo{pages}{2464--2470}.
\newblock
\urldef\tempurl%
\url{https://doi.org/10.24963/ijcai.2017/343}
\showDOI{\tempurl}


\bibitem[\protect\citeauthoryear{Mikolov, Sutskever, Chen, Corrado, and
  Dean}{Mikolov et~al\mbox{.}}{2013}]%
        {DBLP:conf/nips/MikolovSCCD13}
\bibfield{author}{\bibinfo{person}{Tomas Mikolov}, \bibinfo{person}{Ilya
  Sutskever}, \bibinfo{person}{Kai Chen}, \bibinfo{person}{Gregory~S. Corrado},
  {and} \bibinfo{person}{Jeffrey Dean}.} \bibinfo{year}{2013}\natexlab{}.
\newblock \showarticletitle{Distributed Representations of Words and Phrases
  and their Compositionality}. In \bibinfo{booktitle}{\emph{Advances in Neural
  Information Processing Systems 26: 27th Annual Conference on Neural
  Information Processing Systems 2013.}} \bibinfo{pages}{3111--3119}.
\newblock
\urldef\tempurl%
\url{http://papers.nips.cc/paper/5021-distributed-representations-of-words-and-phrases-and-their-compositionality}
\showURL{%
\tempurl}


\bibitem[\protect\citeauthoryear{Mitra, Chaudhari, and Patwardhan}{Mitra
  et~al\mbox{.}}{2014}]%
        {baobaohenshengqi14}
\bibfield{author}{\bibinfo{person}{Sanghamitra Mitra},
  \bibinfo{person}{Nilendra Chaudhari}, {and} \bibinfo{person}{Bipin
  Patwardhan}.} \bibinfo{year}{2014}\natexlab{}.
\newblock \showarticletitle{Leveraging hybrid recommendation system in
  insurance domain}.
\newblock \bibinfo{journal}{\emph{International Journal of Engineering and
  Computer Science}}  \bibinfo{volume}{3} (\bibinfo{date}{Oct.}
  \bibinfo{year}{2014}).
\newblock


\bibitem[\protect\citeauthoryear{Qazi, Fung, Meissner, and Fontes}{Qazi
  et~al\mbox{.}}{2017}]%
        {DBLP:conf/recsys/QaziFMF17}
\bibfield{author}{\bibinfo{person}{Maleeha Qazi}, \bibinfo{person}{Glenn~M.
  Fung}, \bibinfo{person}{Katie~J. Meissner}, {and} \bibinfo{person}{Eduardo~R.
  Fontes}.} \bibinfo{year}{2017}\natexlab{}.
\newblock \showarticletitle{An Insurance Recommendation System Using Bayesian
  Networks}. In \bibinfo{booktitle}{\emph{Proceedings of the Eleventh {ACM}
  Conference on Recommender Systems, RecSys 2017, Como, Italy, August 27-31,
  2017}}. \bibinfo{pages}{274--278}.
\newblock
\urldef\tempurl%
\url{https://doi.org/10.1145/3109859.3109907}
\showDOI{\tempurl}


\bibitem[\protect\citeauthoryear{Rendle, Freudenthaler, Gantner, and
  Schmidt{-}Thieme}{Rendle et~al\mbox{.}}{2012}]%
        {DBLP:journals/corr/abs-1205-2618}
\bibfield{author}{\bibinfo{person}{Steffen Rendle}, \bibinfo{person}{Christoph
  Freudenthaler}, \bibinfo{person}{Zeno Gantner}, {and} \bibinfo{person}{Lars
  Schmidt{-}Thieme}.} \bibinfo{year}{2012}\natexlab{}.
\newblock \showarticletitle{{BPR:} Bayesian Personalized Ranking from Implicit
  Feedback}.
\newblock \bibinfo{journal}{\emph{CoRR}}  \bibinfo{volume}{abs/1205.2618}
  (\bibinfo{year}{2012}).
\newblock
\showeprint[arxiv]{1205.2618}
\urldef\tempurl%
\url{http://arxiv.org/abs/1205.2618}
\showURL{%
\tempurl}


\bibitem[\protect\citeauthoryear{Rokach, Shani, Shapira, Chapnik, and
  Siboni}{Rokach et~al\mbox{.}}{2013}]%
        {DBLP:conf/sac/RokachSSCS13}
\bibfield{author}{\bibinfo{person}{Lior Rokach}, \bibinfo{person}{Guy Shani},
  \bibinfo{person}{Bracha Shapira}, \bibinfo{person}{Eyal Chapnik}, {and}
  \bibinfo{person}{Gali Siboni}.} \bibinfo{year}{2013}\natexlab{}.
\newblock \showarticletitle{Recommending insurance riders}. In
  \bibinfo{booktitle}{\emph{Proceedings of the 28th Annual {ACM} Symposium on
  Applied Computing, {SAC} '13, Coimbra, Portugal, March 18-22, 2013}}.
  \bibinfo{pages}{253--260}.
\newblock
\urldef\tempurl%
\url{https://doi.org/10.1145/2480362.2480417}
\showDOI{\tempurl}


\bibitem[\protect\citeauthoryear{{Scarselli}, {Gori}, {Tsoi}, {Hagenbuchner},
  and {Monfardini}}{{Scarselli} et~al\mbox{.}}{2009}]%
        {4700287}
\bibfield{author}{\bibinfo{person}{F. {Scarselli}}, \bibinfo{person}{M.
  {Gori}}, \bibinfo{person}{A.~C. {Tsoi}}, \bibinfo{person}{M. {Hagenbuchner}},
  {and} \bibinfo{person}{G. {Monfardini}}.} \bibinfo{year}{2009}\natexlab{}.
\newblock \showarticletitle{The Graph Neural Network Model}.
\newblock \bibinfo{journal}{\emph{IEEE Transactions on Neural Networks}}
  \bibinfo{volume}{20}, \bibinfo{number}{1} (\bibinfo{year}{2009}),
  \bibinfo{pages}{61--80}.
\newblock
\urldef\tempurl%
\url{https://doi.org/10.1109/TNN.2008.2005605}
\showDOI{\tempurl}


\bibitem[\protect\citeauthoryear{Schlichtkrull, Kipf, Bloem, van~den Berg,
  Titov, and Welling}{Schlichtkrull et~al\mbox{.}}{2018}]%
        {DBLP:conf/esws/SchlichtkrullKB18}
\bibfield{author}{\bibinfo{person}{Michael~Sejr Schlichtkrull},
  \bibinfo{person}{Thomas~N. Kipf}, \bibinfo{person}{Peter Bloem},
  \bibinfo{person}{Rianne van~den Berg}, \bibinfo{person}{Ivan Titov}, {and}
  \bibinfo{person}{Max Welling}.} \bibinfo{year}{2018}\natexlab{}.
\newblock \showarticletitle{Modeling Relational Data with Graph Convolutional
  Networks}. In \bibinfo{booktitle}{\emph{The Semantic Web - 15th International
  Conference, {ESWC} 2018, Heraklion, Crete, Greece, June 3-7, 2018,
  Proceedings}}. \bibinfo{pages}{593--607}.
\newblock
\urldef\tempurl%
\url{https://doi.org/10.1007/978-3-319-93417-4\_38}
\showDOI{\tempurl}


\bibitem[\protect\citeauthoryear{Sun, Han, Yan, Yu, and Wu}{Sun
  et~al\mbox{.}}{2011}]%
        {DBLP:journals/pvldb/SunHYYW11}
\bibfield{author}{\bibinfo{person}{Yizhou Sun}, \bibinfo{person}{Jiawei Han},
  \bibinfo{person}{Xifeng Yan}, \bibinfo{person}{Philip~S. Yu}, {and}
  \bibinfo{person}{Tianyi Wu}.} \bibinfo{year}{2011}\natexlab{}.
\newblock \showarticletitle{PathSim: Meta Path-Based Top-K Similarity Search in
  Heterogeneous Information Networks}.
\newblock \bibinfo{journal}{\emph{{PVLDB}}} \bibinfo{volume}{4},
  \bibinfo{number}{11} (\bibinfo{year}{2011}), \bibinfo{pages}{992--1003}.
\newblock


\bibitem[\protect\citeauthoryear{Velickovic, Cucurull, Casanova, Romero,
  Li{\`{o}}, and Bengio}{Velickovic et~al\mbox{.}}{2018}]%
        {DBLP:conf/iclr/VelickovicCCRLB18}
\bibfield{author}{\bibinfo{person}{Petar Velickovic}, \bibinfo{person}{Guillem
  Cucurull}, \bibinfo{person}{Arantxa Casanova}, \bibinfo{person}{Adriana
  Romero}, \bibinfo{person}{Pietro Li{\`{o}}}, {and} \bibinfo{person}{Yoshua
  Bengio}.} \bibinfo{year}{2018}\natexlab{}.
\newblock \showarticletitle{Graph Attention Networks}. In
  \bibinfo{booktitle}{\emph{Proceesings of the 6th International Conference on
  Learning Representations, {ICLR} 2018, Vancouver, BC, Canada, April 30 - May
  3, 2018}}.
\newblock
\urldef\tempurl%
\url{https://openreview.net/forum?id=rJXMpikCZ}
\showURL{%
\tempurl}


\bibitem[\protect\citeauthoryear{Wang, Yu, Zheng, Gan, Gai, Ye, Li, Zhou,
  Huang, Ma, Huang, Guo, Zhang, Lin, Zhao, Li, Smola, and Zhang}{Wang
  et~al\mbox{.}}{2019b}]%
        {wang2019dgl}
\bibfield{author}{\bibinfo{person}{Minjie Wang}, \bibinfo{person}{Lingfan Yu},
  \bibinfo{person}{Da Zheng}, \bibinfo{person}{Quan Gan}, \bibinfo{person}{Yu
  Gai}, \bibinfo{person}{Zihao Ye}, \bibinfo{person}{Mufei Li},
  \bibinfo{person}{Jinjing Zhou}, \bibinfo{person}{Qi Huang},
  \bibinfo{person}{Chao Ma}, \bibinfo{person}{Ziyue Huang},
  \bibinfo{person}{Qipeng Guo}, \bibinfo{person}{Hao Zhang},
  \bibinfo{person}{Haibin Lin}, \bibinfo{person}{Junbo Zhao},
  \bibinfo{person}{Jinyang Li}, \bibinfo{person}{Alexander~J Smola}, {and}
  \bibinfo{person}{Zheng Zhang}.} \bibinfo{year}{2019}\natexlab{b}.
\newblock \showarticletitle{Deep Graph Library: Towards Efficient and Scalable
  Deep Learning on Graphs}.
\newblock \bibinfo{journal}{\emph{ICLR Workshop on Representation Learning on
  Graphs and Manifolds}} (\bibinfo{year}{2019}).
\newblock
\urldef\tempurl%
\url{https://arxiv.org/abs/1909.01315}
\showURL{%
\tempurl}


\bibitem[\protect\citeauthoryear{Wang, He, Nie, and Chua}{Wang
  et~al\mbox{.}}{2017}]%
        {DBLP:conf/sigir/Wang0NC17}
\bibfield{author}{\bibinfo{person}{Xiang Wang}, \bibinfo{person}{Xiangnan He},
  \bibinfo{person}{Liqiang Nie}, {and} \bibinfo{person}{Tat{-}Seng Chua}.}
  \bibinfo{year}{2017}\natexlab{}.
\newblock \showarticletitle{Item Silk Road: Recommending Items from Information
  Domains to Social Users}. In \bibinfo{booktitle}{\emph{Proceedings of the
  40th International {ACM} {SIGIR} Conference on Research and Development in
  Information Retrieval, Shinjuku, Tokyo, Japan, August 7-11, 2017}}.
  \bibinfo{pages}{185--194}.
\newblock
\urldef\tempurl%
\url{https://doi.org/10.1145/3077136.3080771}
\showDOI{\tempurl}


\bibitem[\protect\citeauthoryear{Wang, Ji, Shi, Wang, Ye, Cui, and Yu}{Wang
  et~al\mbox{.}}{2019a}]%
        {DBLP:conf/www/WangJSWYCY19}
\bibfield{author}{\bibinfo{person}{Xiao Wang}, \bibinfo{person}{Houye Ji},
  \bibinfo{person}{Chuan Shi}, \bibinfo{person}{Bai Wang},
  \bibinfo{person}{Yanfang Ye}, \bibinfo{person}{Peng Cui}, {and}
  \bibinfo{person}{Philip~S. Yu}.} \bibinfo{year}{2019}\natexlab{a}.
\newblock \showarticletitle{Heterogeneous Graph Attention Network}. In
  \bibinfo{booktitle}{\emph{The World Wide Web Conference, {WWW} 2019, San
  Francisco, CA, USA, May 13-17, 2019}}. \bibinfo{pages}{2022--2032}.
\newblock
\urldef\tempurl%
\url{https://doi.org/10.1145/3308558.3313562}
\showDOI{\tempurl}


\bibitem[\protect\citeauthoryear{Xu, Lian, Han, Li, Xu, and Xie}{Xu
  et~al\mbox{.}}{2019}]%
        {DBLP:conf/cikm/XuLHLX019}
\bibfield{author}{\bibinfo{person}{Fengli Xu}, \bibinfo{person}{Jianxun Lian},
  \bibinfo{person}{Zhenyu Han}, \bibinfo{person}{Yong Li},
  \bibinfo{person}{Yujian Xu}, {and} \bibinfo{person}{Xing Xie}.}
  \bibinfo{year}{2019}\natexlab{}.
\newblock \showarticletitle{Relation-Aware Graph Convolutional Networks for
  Agent-Initiated Social E-Commerce Recommendation}. In
  \bibinfo{booktitle}{\emph{Proceedings of the 28th {ACM} International
  Conference on Information and Knowledge Management, {CIKM} 2019, Beijing,
  China, November 3-7, 2019}}. \bibinfo{pages}{529--538}.
\newblock
\urldef\tempurl%
\url{https://doi.org/10.1145/3357384.3357924}
\showDOI{\tempurl}


\bibitem[\protect\citeauthoryear{Zhang, Song, Huang, Swami, and Chawla}{Zhang
  et~al\mbox{.}}{2019}]%
        {DBLP:conf/kdd/ZhangSHSC19}
\bibfield{author}{\bibinfo{person}{Chuxu Zhang}, \bibinfo{person}{Dongjin
  Song}, \bibinfo{person}{Chao Huang}, \bibinfo{person}{Ananthram Swami}, {and}
  \bibinfo{person}{Nitesh~V. Chawla}.} \bibinfo{year}{2019}\natexlab{}.
\newblock \showarticletitle{Heterogeneous Graph Neural Network}. In
  \bibinfo{booktitle}{\emph{Proceedings of the 25th {ACM} {SIGKDD}
  International Conference on Knowledge Discovery {\&} Data Mining, {KDD} 2019,
  Anchorage, AK, USA, August 4-8, 2019}}. \bibinfo{pages}{793--803}.
\newblock


\end{thebibliography}
\end{document}